\documentclass[rmp,aps,twocolumn,article,nofootinbib,showkeys,showpacs]{revtex4-1} 

\usepackage{amsmath,mathrsfs,amssymb,mathtools} 
\usepackage{amsfonts} 
\usepackage{graphicx}
\usepackage{accents}
\usepackage{pstricks-add}
\definecolor{vert}{rgb}{0,0.65,0}
\usepackage[colorlinks=true,citecolor=blue,urlcolor=red,linkcolor=red]{hyperref}

\hypersetup{
    bookmarks=true,       
    unicode=false,      
    pdftoolbar=true,      
    pdfmenubar=true,      
    pdffitwindow=false,   
    pdfstartview={FitH},   
    pdftitle={On the wonderfulness of Noether's theorems, 100 years later, and Routh reduction},
    pdfauthor={Raphaël Leone},   
    pdfsubject={On the wonderfulness of Noether's theorems, 100 years later, and Routh reduction}, 
}

\usepackage[german,english]{babel}

\newcommand{\sast}{{\scriptscriptstyle *}}

\newcommand\ringring[1]{%
  {
   \mathop{\kern0pt #1}\limits^{
     \vbox to-1.85ex{
       \kern-2ex 
       \hbox to 0pt{\hss\normalfont\kern.1em \r{}\kern-.45em \r{}\hss}%
       \vss 
     }
   }
  }
}

\newcommand*{\rmd}{\ensuremath{\text{d}}}

\let\oldFootnote\footnote
\newcommand\nextToken\relax

\renewcommand\footnote[1]{%
    \oldFootnote{#1}\futurelet\nextToken\isFootnote}

\newcommand\isFootnote{%
    \ifx\footnote\nextToken\textsuperscript{,}\fi}
    
    \makeatletter
\let\Hy@linktoc\Hy@linktoc@none
\makeatother

\makeatletter
\def\l@subsubsection#1#2{}
\makeatother

\begin{document}

\title{On the wonderfulness of Noether's theorems, 100 years later, \\ and Routh reduction}

\author{Rapha\"el Leone}
\email{raphael.leone@univ-lorraine.fr} 
\affiliation{-- Laboratoire de Physique et Chimie Th\'eoriques (UMR CNRS 7019), F-54506 Vand\oe uvre-l\`es-Nancy Cedex,}

\affiliation{-- Laboratoire d'Histoire des Sciences et de Philosophie -- Archives Henri Poincar\'e (UMR CNRS 7117), F-54501 Nancy Cedex, \\ Universit\'e de Lorraine, France}

\date{\today}

\begin{abstract}
\textbf{\sffamily Abstract.} This paper is written in honour of the centenary of Emmy Amalie Noether's famous article entitled \emph{Invariante Variationsprobleme}. It firstly aims to give an exposition of what we believe to be the most significant and elegant issues regarding her theorems, through the lens of classical mechanics. Despite the limitation to this field, we try to illustrate the key ideas of her work in a rather complete and pedagogical manner which, we hope, presents some original aspects. The notion of symmetry coming naturally with the idea of simplification, the last part is devoted to the interplay between Noether point symmetries and the reduction procedure introduced by Edward John Routh in 1877. 
\begin{flushright}
\textsl{Le temps d'apprendre \`a vivre il est d\'ej\`a trop tard}\\
Louis Aragon
\end{flushright}
\end{abstract}



\maketitle 

\tableofcontents

\section{Introduction}

The title of the present article, in its first part, is borrowed from the relatively recent book of Neuenschwander \cite{Neuenschwander}. Indeed, we can only endorse the adjective \emph{wonderful} which is certainly the first to come to our mind concerning Noether's theorems \cite{Noether}. They establish a profound and elegant correspondence, in variational problems, between symmetries and conservation laws or identities, depending on whether the symmetry group is finite (first theorem) or not (second theorem). Because variational formulations are ubiquitous in physics, if not universal in Nature, one easily understands the fascination they arouse in the mind of physicists\footnote{Neunschwander's book \cite{Neuenschwander} is one such example. Notwithstanding its remarkable aesthetic and pedagogical dimensions, we believe that the connection it makes between the first theorem of Noether in classical mechanics and the theory of adiabatic invariants is overestimated [for a proper introduction to this issue, see e.g. chapter 9 of \textcite{Boccaletti}].} and also why they are considerably commented since more than half a century \cite{Kosmann}.

\begin{figure}
\centering
\includegraphics[scale=.35]{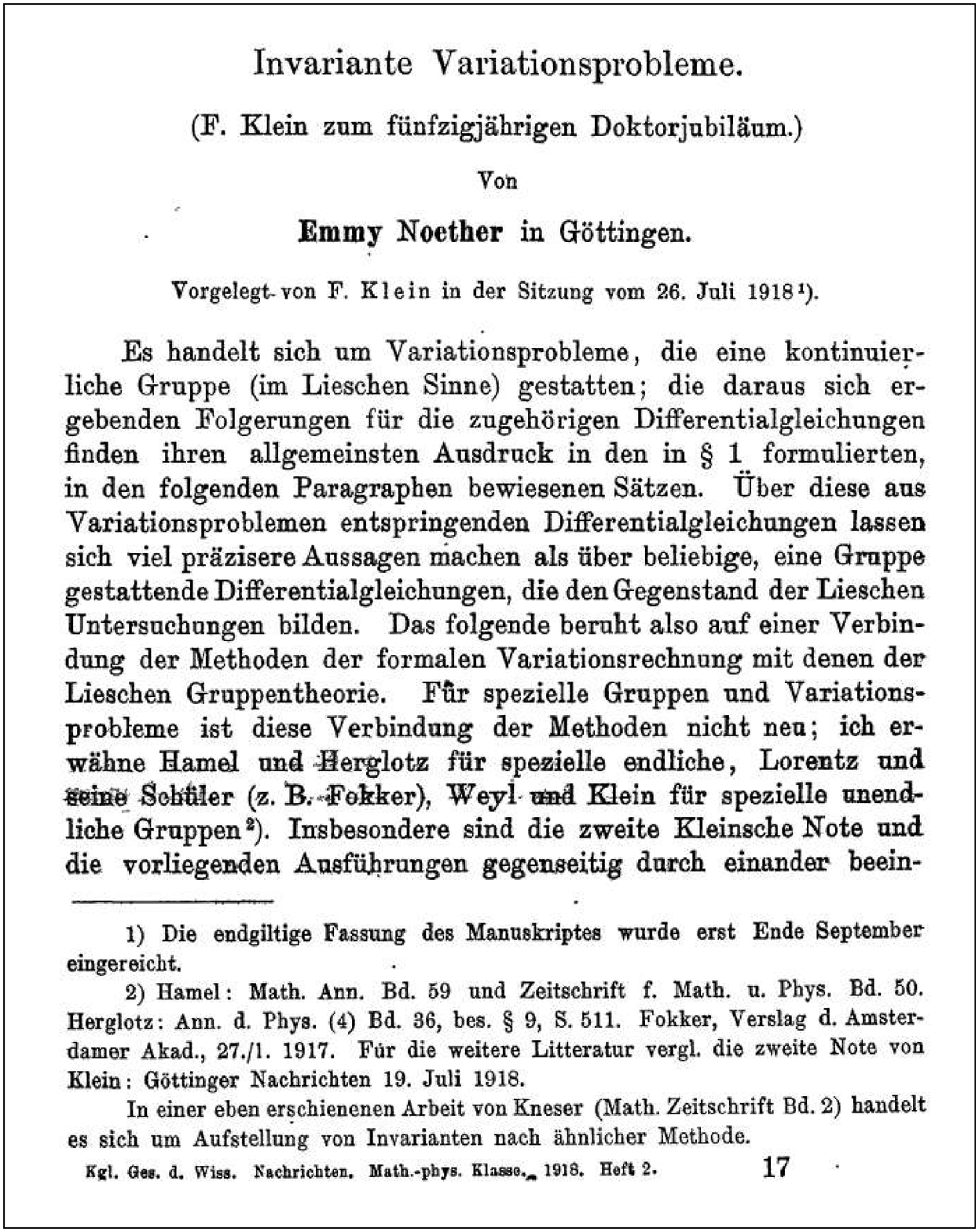}
\caption{A capture of the front page of Noether's paper.}
\end{figure}

The original paper of Emmy Amalie Noether, presented by Klein, appeared in 1918, at a time when mathematicians raised the issue of the mathematical foundations of General Relativity. G\"ottingen was the center of this activity, under the impetus of Klein and Hibert. Invited by these two scientists to join them, she was in particular wondering about what becomes of the conservation laws of energy and momenta, which are considered as cornerstones of physics, and which had been well characterized in classical mechanics since Lagrange \cite{Lagrange} and Jacobi \cite{Jacobi1866} through the absence of a variable in a dynamical function now known as Lagrangian. Such an independence is the manifestation of a symmetry under a displacement along the direction of the involved variable and is often presented, with good reason, as a very first encounter with Noether's theory.

This introduction is not the place to give a faithful account of the circumstances which led Noether to state her famous theorems; it would take us too far from the scope of the present paper.\footnote{On the historical and bibliographical aspects, the interested reader may consult \textcite{Dick}, \textcite{Pais}, \textcite{Kastrup}, \textcite{Byers1996,Byers1999}, and \textcite{Kosmann}.} Let us only give some elements of the context. In 1911, Herglotz~\cite{Herglotz} had already exhibited the relationship between the Poincar\'e group and the ten conservation laws of special relativity by using variational techniques. Taking Herglotz's work into consideration, Klein thought that it was linked to Lie's theory\footnote{Lie was mainly interested in the questions around the invariances of differential equations but left untouched the issue of the invariances of variational principles \cite{Rowe2002}.} and suggested to Engel --- its former student who became a close collaborator of Lie\footnote{Engel was notably the co-author of Lie's celebrated \textit{Theorie der Transformationsgruppen} \cite{Engel_Lie} in three volumes.} --- the idea of analysing the interplay between Galilean symmetries, in Lie's sense, and the classical conservation laws. This was the object of a note written by Engel\footnote{It is striking that the method used by Engel is exactly what is nowadays known as the Hamiltonian counterpart of the first theorem of Noether in classical mechanics. It is the occasion for us to indicate that we will not be concerned with the Hamiltonian aspect of the theory in this paper. Not that this is not an interesting issue on its own but we want to remain close to the original formulation. More generally, with any reformulation of mechanics will be assigned its version of Noether's first theorem.} and presented by Klein in 1916 \cite{Engel}. As mentioned in the abstract of her paper, Noether had, in a sense, conjugated these two approaches (variational techniques and Lie's analysis of invariances) while providing a high degree of generality.

The crucial point of Noether's work was the discovery of a profound dichotomy, in terms of their consequences, between two kinds of symmetry groups leaving invariant the functional integral of a variational problem: ̀`continuous finite' (\textit{endlische kontinuierlische}) groups on the one hand and `continuous infinite' (\textit{unendlische kontinuierlische}) groups on the other. She showed that the former give rise to `divergence relations' (\textit{Divergenzrelationen}) and then to `proper' (\textit{eigentliche}) conservation laws, in contrast with the latter which lead to identities (\textit{Abh\"angigkeiten}) at the source of `improper' (\textit{uneigentliche}) conservation laws. In Noether's terminology, a proper conservation law is the expression that a quantity has a divergence only vanishing on-shell (i.e. dynamically) whereas an improper one has a divergence even vanishing off-shell (i.e. identically).\footnote{Bergmann \cite{Bergmann1949,Bergmann_Schiller} introduced the terms `weak' and `strong' instead of `proper' and `improper', respectively.} Divergence relations are, in a sense, constraints on the dynamics whereas identities constraint the theory itself. The latter thus carry the most profound physical meaning and their existence is the \textit{quid pro quo} of the wide arbitrariness lying in the invariance under an infinite group.

Moreover, she established a third theorem which states that a finite symmetry group generates improper conservation laws if and only if it is a subgroup of an infinite symmetry group. These conclusions allowed Noether to prove an assertion of Hilbert whose elucidation was an express request of Klein. The assertion at issue was the conjecture whereby conservation laws of energy and momenta in the space-time of General Relativity were fundamentally improper. Here is how it appeared, in an article of Klein \cite{Klein_Hilbert} which included extracts of letters exchanged between him and Hilbert:\footnote{\textit{\glqq Mit Ihren Ausf\"uhrungen \"uber den Energiesatz stimme ich sachlich v\"ollig \"uberein: Emmy Noether, deren H\"ulfe ich zur Kl\"arung derartiger analytischer meinen Energiesatz betreffenden Fragen vor mehr als Jahresfrist anrief, fand damals, da\ss die von mir aufgestellten Energiekomponenten --- ebenso wie die Einsteinschen --- formal mittelst der Lagrangeschen Differentialgleichungen [\dots] in meiner ersten Mitteilung in Ausdr\"ucke verwandelt werden k\"onnen, deren Divergenz identisch d. h. ohne Benutzung der Lagrangeschen Gleichungen [\dots] verschwindet. Da andererseits die Energiegleichungen der klassischen Mechanik, der Elastizit\"atstheorie und Elektrodynamik nur als Folge der Lagrangeschen Differentialgleichungen des Problems erf\"ullt sind, so ist es gerechtfertigt, wenn Sie deswegen in meinen Energiegleichungen nicht das Analogon zu denen jener Theorien erblicken. Freilich behaupte ich dann, da\ss f\"ur die allgemeine Relativit\"at, d. h. im Falle der allgemeinen Invarianz der Hamiltonschen Funktion, Energiegleichungen, die in ihrem Sinne den Energiegleichungen der orthogonalinvarianten Theorien entsprechen, \"uberhaupt nicht existieren; ja ich m\"ochte diesen Umstand sogar als ein charakteristisches Merkmal der allgemeinen Relativit\"atstheorie bezeichnen. F\"ur meine Behauptung w\"are der mathematische Beweis erbringbar.\grqq}}

\begin{quote}
I fully agree in fact with your statements on the energy theorems: Emmy Noether, on whom I have called for assistance more than a year ago to clarify this type of analytical questions concerning my energy theorem, found at that time that the energy components which I had proposed --- as well as those of Einstein --- could be formally transformed, using the Lagrange differential equations [\dots] of my first note, into expressions whose divergence vanishes identically, i.e. without using the Lagrange equations [\dots]. On the other hand, the energy equations of classical mechanics, the theory of elasticity, and electrodynamics, result only from the verification of the Lagrangian differential equations, so you are justified in seeing in my energy equations not the analogous to those of these theories. It naturally brings me to claim that, for the General Relativity theory, i.e. in the case of the general invariance of the Hamiltonian function,\footnote{The `Hamiltonian function' is actually the Lagrangian present in the Hamilton principle.} the energy equations, which correspond to your views of them regarding orthogonally invariant theories, do not exist at all; I would even regard this fact as a characteristic feature of the General Relativity theory. A mathematical proof of my assertion should be realisable.
\end{quote}

By `orthogonally invariant' theory should be understood a theory whose stage is a world of finite dimension in which is anchored an immutable metric structure --- independent of any other considerations --- admitting a symmetry group. Its elements are the isometries of the space, that is, the transformations leaving the theory `orthogonally invariant'. In special relativity, for example, the isometry group of Minkowski spacetime is the ten-dimensional Poincar\'e group which is linked to the ten proper conservation laws of that theory. More generally, symmetries of the space are exactly the transformations leaving invariant all its immutable background properties given \textit{ab initio}. If one imagines that the latter are encoded in a set of fields, the symmetries will be the transformations leaving all these entities invariant. But, General Relativity is \emph{background independent}; there are no such fields, not even the metric which acquires the status of a dynamical field. Under these circumstances, any transformation is automatically a symmetry and the symmetry group becomes infinite: the theory is therefore \emph{generally invariant} (or \emph{diffeomorphism invariant}\footnote{One can find in \textcite{Pooley2017} an interesting discussion about the interplays between the distinct concepts of general covariance, diffeomorphism invariance, and background (in)dependence.}). 

General invariance and her third theorem --- which stands opportunely in her last section (\textit{Eine Hilbertsche Behauptung}) devoted to Hilbert's assertion --- are all that is needed to achieve her aims. Noether starts by fixing an arbitrary system of coordinates and considers the associated finite `group of translations'\footnote{The quotations marks are presents in Noether's paper. Indeed, it is an abusive use of this term of usual geometry consisting in `doing as if we were dealing with Cartesian coordinates'. As written in \textcite{EIH1938}, `we use the Euclidean nomenclature merely because it is apt and convenient'.} (\textit{\glqq Verschiebungsgruppe\grqq}) along these coordinates. It yields as many `energy relations'\footnote{The quotation marks are present in Noether's paper.} (\textit{\glqq Energierelationen\grqq}) --- the energy equations of Hilbert --- as the number of dimensions of the space. Since this arbitrary group of translations is a finite subgroup of the infinite symmetry group of all the continuous transformations, the energy equations are necessarily improper by her third theorem. Alternatively stated, the energy equations, as understood in the proper sense characteristic of `orthogonally invariant' theories, do not exist in General Relativity (and even, of course, in any generally invariant theory admitting a variational formulation). 

One must actually say that the use of the word `symmetry' is anachronistic. We never encounter it under Noether's pen, only the term `invariance' (\textit{Invarianz}). Indeed, her work is clearly rooted in the wide `theory of invariants'\footnote{Roughly speaking, this theory aims to characterize the operations leaving invariant various mathematical objects in the sense that they retain their form, and to analyse the consequences of these invariances. Obviously, it requires the prior definitions of what is meant by `operation' and `form' but one evidently recognizes in this general picture the concept of symmetry.} which took off in the middle of the nineteenth century\footnote{As is pointed out in \textcite{Kosmann}, H. Weyl \cite{Weylclassical} saw the starting point of the theory of invariants in a memoir presented by Cayley in 1846 \cite{Cayley}. It is actually a slightly extended version (in french) of two preceding papers of Cayley \cite{Cayley1845,Cayley1846}.} and reached a climax with Lie's works on differential invariants through his fecund concept of continuous transformation groups (\textit{continuierlichen Transformationsgruppen}).  As a matter of fact, Noether repeatedly refers to Lie. However, one may be surprised at first glance by the complete absence, in her paper, of the geometric pictures which infused in Lie's works \cite{Lie1891}. She also quite naturally mentions the name of Klein at many places, evokes the Erlangen program, but only in its group aspect,\footnote{Her final note is precisely the recognition of Klein's ideas regarding the true nature of the concept of relativity in terms of invariance with respect to a group (\textit{Hiermit ist wieder die Richtigkeit einer Bemerkung von Klein best\"atigt, da\ss{} die in der Physik \"ubliche Bezeichnung \glqq Relativit\"at\grqq{} zu ersetzen sei durch \glqq Invarianz relativ zu einer Gruppe\grqq}).} without further exploration of the geometric one in spite of its profound signification in General Relativity, that theory at the origin of her research.\footnote{The only detailed discussion in relation with General Relativity is to be found in her twentieth note where she illustrates the fact that a symmetry of a functional integral can be adapted to an equivalent one if a divergence is added to the Lagrangian density. (This statement will be subsequently simplified after the introduction of the concept of invariance up to a divergence.) With Einstein's `$\Gamma-\Gamma$' action \cite{Einstein1916} in mind, she indeed considers Hilbert's action to which she adds a surface term.} Notably, the fact that Noether did not pursue the geometrical significance of the four identities arising from the general invariance speaks for itself.\footnote{These identities --- one for each arbitrariness in the choice of a coordinate --- were later recognized as the contracted Bianchi ones. Hilbert had already derived them in 1915 \cite{Hilbert1915} in a somewhat `convoluted' manner which was not understood at the time and which keeps modern historians busy \cite{Sauer1999,Renn2007}. One can find in \textcite{Rowe2002} an interesting discussion on the `memory loss' of G\"ottingen circles about the Italian differential geometry which explains why they struggled with these identities and also why the latter were rediscovered a certain number of times. The connection with the old works of Bianchi, Padova, and Ricci, was brought to light by Schouten and Struik in 1924 \cite{Schouten-Struik}. Eddington could not be aware of this article when he worked on the second edition of his celebrated book entitled \textit{The Mathematical Theory of Relativity} \cite{Eddington1924} which appeared the same year, and where the unnamed `four identities' are said to constitute the `fundamental theorem of mechanics'. Eddington wrote about them:
\begin{quote}
I think it should be possible to prove [\dots] by geometrical reasoning [\dots]. But I have not been able to construct a geometrical proof and must content myself with a clumsy analytical verification.
\end{quote}
In his no less famous \textit{Space, Time and Gravitation} \cite{Eddington1920} which was published four years earlier, one can also find in a note:
\begin{quote}
I doubt whether anyone has performed the laborious task of verifying these identities by straightforward algebra.
\end{quote}
Much later, the term `Bianchi identities' was introduced in \textcite{Bergmann_Schiller} to name the identities arising from the general invariance, whatever the theory be. Trautman called them `generalized Bianchi identities' in his illuminating articles on the subject \cite{Trautman1963,Trautman1967,Trautman_Witten}, and Anderson used the term `Bianchi-type identities' \cite{Anderson}.}\footnote{In addition, Klein derived from the general invariance four sets of identities, 140 in all \cite{Klein1918}.} The reason for this `lack of geometry' is certainly the undeniable high degree of generality introduced by Noether who proceeded in following an abstract\footnote{Let us recall that Noether is recognized by the mathematical community as a great algebraist above all \cite{Tent}. Today, any undergraduate student in mathematics knows what is a \emph{Noetheran ring} [this term was coined in 1943 by Chevalley \cite{Gilbert1981}].} and rather formal approach, far from any well-characterized geometric basis at that time. Hence, her work did clearly not fall in the domain of geometry but in that of the calculus of variation, a field in which G\"ottingen circles had a considerable level of expertise\footnote{More generally, German mathematics was undoubtedly at the forefront of the variational calculus. To be convinced of this, it suffices to read the preface of Bolza's monograph on the subject \cite{Bolza1904}.} \cite{Rowe2002}. 

The first explicit application of Noether's theorems appeared three years later, in an article of Bessel-Hagen \cite{Bessel-Hagen} who used them to study Galilean invariance in classical mechanics as well as the conformal invariance in electrodynamics. This work was (again\footnote{As Kastrup \cite{Kastrup}, we cannot resist to quote an excerpt of \textcite{Reid1970}:
\begin{quote}
Also, with age, Klein was becoming more olympian. A favorite joke among the students was the following: In G\"ottingen there are two kinds of mathematicians, those who do what they want and not what Klein wants --- and those who do what Klein wants and not what they want. Klein is not either kind. Therefore, Klein is not a mathematician.
\end{quote}
We recall that Klein presented Noether's paper while he was entering his 70's. He died in 1925.}) proposed by Klein and received some support from Noether \cite{Rowe1999}. In particular, she communicated to Bessel-Hagen\footnote{Bessel-Hagen wrote: \textit{\glqq\textit{Ich verdanke diese einer m\"undlichen Mitteilung von Fra\"ulein Emmy Noether}\grqq}.} the idea of naturally generalizing her theory through the introduction of the concept of invariance `up to a divergence' (\textit{bis auf eine Divergenz}). The motivation behind this refinement is to be found in either the analysis of Galilean boosts, or the general invariance issue looked through Einstein's `$\Gamma-\Gamma$' action \cite{Einstein1916} which is built on a non scalar density and thus do not admit the symmetry group of general invariance in the original sense of Noether \cite{Brown2002}.

Shortly afterwards, Weitzenb{\"o}ck\footnote{The hidden message contained in the introduction of Weitzenb{\"o}ck's book (\textit{Nieder mit den Franzosen}) is a proof (if any were required) that there is no incompatibility between being a recognized scientist and a prize idiot.} \cite{Weitzenbock} and, above all, Courant and Hilbert\footnote{The part devoted to Noether's theorems was enlarged in the second edition appeared in 1931.} \cite{Courant-Hilbert}, were the first to disseminate some aspects of Noether's results through their textbooks. However, the book of Courant and Hilbert truly became a classic of mathematical physics only after the publication of the English version, in 1953. It is maybe one of the reasons why it seems that they were not widely spread in the scientists community during the three decades following their publication \cite{Kosmann}, \textit{a fortiori} in non German-speaking circles. In fact, they remained almost confined to G{\"o}ttingen circles \cite{Kastrup}. One other reason is perhaps the fact that the usual invariances of physical theories and their association with conservation laws acquired a certain status of `common knowledge' \cite{Salisbury2012} which did not call for all the generality contained in Noether's work, whereas there were only a few number of works in relation with Noether's one. On this aspect, Bergmann provides an interesting example. In his article of 1949 --- in which is examined the conservation laws and identities arising in generally invariant classical field theories by using variational techniques \textit{\`a la} Noether ---, he did not mention Noether, nor any other German scientist.\footnote{We point out the use of the symbol $\bar\delta$ to denote the `variation of the field variables as functions of their arguments' in Bergmann's article, whereas $\delta$ symbolizes the total variation. It is exactly the same convention than in Noether's article as well as in Courant and Hilbert's book. Actually, the $\bar\delta$ symbol which operates the `substantial variation' \cite{Rosenfeld} coincides with the $\delta$ commonly used in Hamilton's principle. This is maybe the reason why Bessel-Hagen or Weitzenb\"ock, among others, preferred reserve $\delta$ to the substantial variation.} This absence of reference to Noether is somehow astonishing from a native German physicist, and a former research assistant of Einstein who had shown himself impressed by Noether's work in his correspondence with Klein and Hilbert during 1918 \cite{Kosmann}. In fact, among Bergmann's articles which were published in the fifties on issues related to the general invariance, there is only a single mention of Noether: it concerned her third theorem \cite{Bergmann1953}. 

Beyond considerations about the dissemination of knowledge in general, and the possible reasons evoked above, we can reasonably think that Noether's `virtuosity' in the calculus of variation was not such as to contribute to spreading her work. Indeed, she seemed so comfortable with abstract variational techniques that her article certainly lacked of explanations for the `ordinary physicist'. Of course, Bessel-Hagen expanded the physical contents but it could be argued that it was not sufficient. There is a good degree of consensus in considering that Hill's pedagogical paper \cite{Hill}, which appeared two years before the English version of Courant and Hilbert's book, was the first to popularize Noether's first theorem \cite{Kosmann,Olver,Logan}, although he did not name it in this way but referred to the collective contribution of Klein, Noether, and Bessel-Hagen. However, it is sometimes reproached to Hill \cite{Olver,Kosmann} the fact that he only raised the issue of the first theorem, and --- to make things worse --- in the `simplest' case of Lagrangian densities of the first order. Olver writes that Hill caused the belief whereby `this was all Noether had proved on the subject', while Kosmann-Schwarzbach even goes so far as to say that Hill was a `well-intentioned culprit' who `completely denatured her results'. These are snap judgements. Nowhere did he claim to give a faithful exposition of Noether's theory (incidentally, we recall that he did not mention her name alone), and the second theorem of Noether was out of the scope (all the more so he did not develop the group aspect). Moreover, the restriction to first order Lagrangians was welcome for pedagogical reasons\footnote{Hill's pedagogical intentions are unambiguous. They are presented in the last paragraph of his introduction, as follows: \begin{quote}
Despite the fundamental importance of this theory there seems to be no readily available account of it which is adapted to the needs of the student of mathematical physics, while the original papers are not readily accessible. It is the object of the present discussion to provide a simplified account of the theory which it is hoped will be of assistance to the reader in gaining an idea of the concepts underlying this important problem. In order to clarify the relationship of the equations of motion and the conservation theorems, as they follow from the variational principle, we shall give a systematic review of the derivations of both sets of equations.
\end{quote}
} and he did not forget to stipulate that the theory was generalizable to any order.\footnote{He wrote:
\begin{quote}
While this restriction is adequate to cover the cases normally met with in physical problems, the mathematical theory can be generalized to
include derivatives of any desired order.
\end{quote}
} It allowed him to detail the different steps towards the main result, while remaining extremely clear in his explanations. In sum, Hill did what is expected of a pedagogical article: he brought to light an important aspect of a theory and rendered it accessible to students, without suggesting that it was all that can be said on the subject. If the alleged belief claimed by Olver is true, Hill cannot be held responsible of it.\footnote{However, we do not pretend that Hill's paper is beyond criticism. One can regret, with \textcite{Kosmann}, that Hill restricted himself to (some) ̀`classical symmetries' despite the fact that it was not necessary or, with \textcite{Brading2003}, that the divergence term was only assumed to be of the same order than the Lagrangian density.}

The fact that Hill did not discuss Noether's second theorem can be possibly explained by its lack of physical applications at that time. We recall that this theorem proved to be central in Gauge Theories which truly got off the ground three years later.\footnote{Although the gauge concept was born long before \cite{ORaifeartaigh1997}, it was awaiting a convincing physical realization (other than electromagnetism).} It was launched by the celebrated paper of Yang and Mills \cite{YangMills} on the isotopic spin and was elevated as an unifying principle by Utiyama \cite{Utiyama}. The latter proposed, on a variational basis, to interpret any interaction as the net result of an enlargement of an initially finite symmetry group to an infinite one.\footnote{Today one would say that the initial global symmetry is rendered local \cite{Ryder}, although the words `global' and `local' have a specific meaning in physics that does not match the mathematical definition of these terms.} Consequently, the initial proper conservation laws become improper and the theory acquires fundamental constraints which make it a `gauge theory of the Yang-Mills type'. We remark \textit{en passant} that, although Utiyama's work is closely related to Noether's one, it nevertheless contains no mention of her.\footnote{However, he refers to Rosenfeld \cite{Rosenfeld} who explicitly mentioned Noether. Rosenfeld, to a certain extent, anticipated Bergmann's works on generally invariant theories in view of their quantization as well as that of Utiyama on Gauge Theories \cite{Salisbury}. In particular, he also derived, \textit{\`a la} Klein, the set of identities stemming from the variational invariance under an infinite group. Utiyama applied the same procedure.} One can say that a Yang-Mills theory is the perfect physical realization of Noether's ideas: her first theorem applies in the absence of interaction whereas the second does in their presence. In Yang-Mills theories --- which are the cornerstones of the Standard Model --- the symmetry acquires a creative and organizational strength. Today, a textbook on field theories which would not allude to these two theorems seems something inconceivable.

Once Gauge Theories made their entrance, all the physics started to become reconsidered on symmetry basis; Noether's theorems undoubtedly gained a growing interest, and the literature on the subject was on the increase. In a certain extent, the modernity in physics is discriminated by Noether's second theorem. To the question `what is modern physics ?' can be answered `the physics in which Noether's second theorem plays a role'.\footnote{It should be added: `other than a marginal one'. As will be reviewed in the present paper, one can make this theorem manifest even in classical mechanics through an extended formulation due to Weierstrass, where a parametrization freedom is introduced. In fact, this can be done for any `non modern' theory in the sense given above \cite{Kurchar,Westman}. Rephrasing \textcite{Pooley2017}, one could rather say that the modernity in physics lies in the absence of a formulation where Noether's second theorem would not play a prominent role.} The most well-established modern theories are obviously General Relativity and Yang-Mills theory (underlying the Standard Model). The second theorem of Noether is the source of strong analogies between the frameworks of General Relativity on the one hand, and Yang-Mills theories on the other. They both share the essential existence of arbitrariness leading to fundamental constraints on the theories and on an unavoidable underdetermination in their dynamical equations.\footnote{In General Relativity, this underdetermination caused profound troubles, notably in Einstein's mind [a good reference on this point, in particular about his \textit{a priori} paradoxal `hole argument', is \textcite{NortonHole}]. Indeed, it implies the non-uniqueness of the solutions of the equations of motion. The problem was resolved by the recognition of the too much `degree of reality' assigned with the coordinates, incompatible with the active view of the general invariance. Coordinates, as well as the points they represent, must fundamentally be seen as \textit{insignificant entities} \cite{Westman}. Only with the relation between points can be assigned a physical meaning (see the `point-coincidence' argument of Einstein in Norton's paper) and `gauge conditions' on coordinates are necessary to recover a determinism in the sense of Cauchy-Kowalewski \cite{Anderson}.} It must be noticed that they were already seen and fruitfully developed by Weyl \cite{Weyl1929} --- the recognized father of the gauge idea \cite{Weyl1918} --- at his day, when electromagnetism was the only well-established gauged interaction. In particular, the vocabulary of Gauge Theories is commonly used in General Relativity: general invariance is sometimes presented as its gauge group whereas conditions on (or choices of) coordinates are often called gauge conditions (or fixings). However, there also remain strong differences stemming from the nature of the infinite symmetry groups: gauge invariance is an \emph{internal} symmetry while the general invariance is \emph{external}, in the sense that the latter concerns the surrounding space-time itself and not extra degrees of freedom (describing properties of the matter). This discrepancy has important mathematical consequences which leave aside the General Relativity from the Standard Model and has naturally motivated constant research since the sixties in order to genuinely `gauge' the gravitation\footnote{In such a theory, the gauge group can definitely not be the group of diffeomorphisms encoding the general invariance.} in the light of the other interactions \cite{Blagojevic,Hehl}. This issue --- which is in many aspects comparable to the erstwhile attempts of providing unified field theories \cite{Tonnelat} --- is still open.

As we have just seen, discussing Noether's theorems can brings us to cover all the most fundamental questions of modern physics, while it is commonplace to mention the growing degree of sophistication of physical theories. Since the sixties, their formulations have become more and more intrinsic, that is, their statements appealed less and less to arbitrariness (of choosing a coordinate, a gauge, etc.). For this program to be sustainable, it necessitates a rigorous characterization of the structures where the significant mathematical objects live. The universal language turned out to be that of fiber bundles.\footnote{Today, a `bundlization' of a theory can be, in some extent, taken as a synonym for its `geometrization' whereas we believe that 'geometrizing' is close to `understanding'.} The main contributor to the appearance of this mathematical apparatus in physics was certainly Trautman. He notably exposed to physicists the bundle structures associated with Gauge Theories and General Relativity, and pointed out their differences \cite{Trautman1970,Trautman1979,Trautman1980,Trautman1981}. More interestingly for the purpose of the present paper, it seems that he was the first to give a formulation of a class of Noether symmetries by using the framework of the jet bundles \cite{Trautman1967} which suitably `geometrizes' the theory of differential equations. Now, rigorous mathematical expositions on the subject of Noether (and Lie) symmetries have reached a high degree of sophistication \cite{Olver,Sardanashvily} and Noether's theory can serenely fall in the domain of (differential) geometry.

In parallel to the aforementioned growth of sophistication, basic applications and expositions of Noether's theorems have never ceased to appear in the literature. This is especially true in the realm of classical mechanics, for obvious pedagogical reasons, and also because that domain is in the common culture of physicists (and some engineers) from a long time. Since the first theorem is by far the most relevant in classical mechanics (or, more generally, in the non modern theories according to the characterization given above), it is frequently taken as \textit{the} theorem of Noether\footnote{Interestingly, the plural in the German editions of Courant and Hilbert's book (\textit{Die S\"atze von E. Noether}) became `Noether's theorem' in the English version. Even in the second edition of Bergmann's \textit{Introduction to the Theory of Relativity} \cite{Bergmannbook}, one can find a paragraph devoted to Noether's theory entitled `Noether's theorem' although it contains a discussion on infinite symmetry groups.} (to the great displeasure of Noether's admirers). In classical mechanics textbooks, this theorem is often presented for the sake of elegance since the conservation laws (in time) thereby derived are generally already known by the students; not forgetting that the prominent \textit{r\^ole} played by the symmetry in modern physics naturally motivates to familiarize them with this concept, whenever possible.\footnote{It is now not unusual to encounter textbooks structured around the key concept of symmetry. One such example is \textcite{Doughty}. Let us mention another and surprising book, \textcite{Sudarshan}, in which the symmetry plays an important role but where no reference to Noether is made.} In our objective to write a paper in the honour of the centenary of Noether's one which would be both accessible for most readers and almost self-contained, we decided to restrict ourself to this field. However, we do believe that it allows to illustrate the key ideas of Noether's works. Furthermore, it will give us the opportunity to make a natural link with an old recipe introduced by Routh \cite{Routh1877} to decrease the number of degrees of freedom when ignorable coordinates are present. Since, as will be reviewed, ignorable coordinates are the manifestations of some symmetries, Routh procedure will ideally finalise our work on the expected aspect of a symmetry: its capability of generating a simplification.

The paper is organized as follows. Section~\ref{sec:transformations} starts with a `modern' presentation of the notion of continuous point transformations in the space of events and puts the emphasis on the underlying geometry. Then, the prolongation of their action on kinematics and on scalar quantities is reviewed. The stage having been set, we define continuous point symmetries before exploring their meanings from different viewpoints (active versus passive) and their capacity of reducing by one the number of variables through the introduction of adapted coordinates. 

Section~\ref{sec:Noether} is devoted to Noether's theorem \textit{per se} in classical mechanics by restricting its range of application to the most meaningful point symmetries (some considerations on the generalized symmetries, which are essential to the converse of that theorem, are reported in the appendix). Some useful characterizations of Noether point symmetries (NPS) are then established and their associated first integral are obviously derived. This part of the study ends with invariance issues, regarding the couple formed by a symmetry and its first integral, under Lagrangian gauge and coordinate transformations. 

Section~\ref{sec:applications} develops three applications for the purpose of exploring different aspects of the theory. The first application aims to characterize the NPS admissible by the most standard Lagrangian form encountered in classical mechanics. It is the occasion to exploit the advantages of the form invariance. In the second application, we determine all the NPS admitted by `natural problems' in the one dimensional case, through the use of adapted coordinates. The last one deals with the parametrization invariance as a manifestation of the second theorem in classical mechanics, an example given by Noether herself. 

The final section~\ref{sec:Routh} is devoted to the interplay between NPS and the Routh reduction procedure. In its first part, that procedure is reviewed and its role of bridge between variational principles as well as its connection with an old theorem of Whittaker are clarified. Then, in its last part, we explain in an accessible manner why abelian groups of NPS are the only one to which Routh reduction applies.

\section{Continuous point transformations and symmetries}\label{sec:transformations}

\subsection{Generalities on continuous point transformations}

Let us consider a mechanical system whose configuration space is, as usual, a smooth\footnote{The adjective `smooth' refers to some $\mathscr C^r$ property with $r\geqslant 1$. We recall that a function (or mapping) is said to be $\mathscr C^r$ if its $r$ first derivatives exist and are continuous. To put it simple, a manifold is $\mathscr C^r$ if one can use everywhere systems of coordinates which transform between themselves in a $\mathscr C^r$ way. In what follows, the manifold $\mathcal M$ will be assumed as smooth as required by the statements under consideration.} manifold $\mathcal M$ of finite dimension $n$. Since we are dealing with Newtonian mechanics, it is assumed that an absolute timeline $\mathcal T$ exists, whose points are the `positions in time'. It is in itself a smooth manifold of dimension 1 admitting a global time coordinate $t$. Taking its Cartesian product with $\mathcal M$ yields a smooth manifold $\mathcal E=\mathcal T\times\mathcal M$ of dimension $n+1$ known as the \emph{extended configuration space} (or \emph{space of events}). Any point $p$ of $\mathcal E$ identifies with a couple $(t,q)$ where $t$ is a position in time and $q$ in $\mathcal M$. Everywhere, $\mathcal E$ is locally describable by means of extended coordinate systems $\{t,q^i\}$ where $q^i$ ($i=1,\dots,n$) are coordinates in $\mathcal M$.

\begin{figure}
\centering
\psset{xunit=.6cm,yunit=.6cm,algebraic=true,dotstyle=o,dotsize=3pt 0,linewidth=0.5pt,arrowsize=3pt 2,arrowinset=0.25}
\begin{pspicture*}(0.05,0.1)(14.4,8.95)
\parametricplot[linewidth=.3pt]{1.7093114184455844}{2.696464557035488}{1*4.71*cos(t)+0*4.71*sin(t)+4.94|0*4.71*cos(t)+1*4.71*sin(t)+-0.68}
\parametricplot[linewidth=.3pt]{1.651358282808984}{2.6743618432795206}{1*3.5*cos(t)+0*3.5*sin(t)+4.63|0*3.5*cos(t)+1*3.5*sin(t)+-0.94}
\psline{->}(0.08,0.38)(5.03,0.4)
\psline{->}(9.4,0.38)(14.35,0.4)
\rput{18.19}(7.52,6.81){\psellipse(0,0)(3.17,1.91)}
\psline[linewidth=.3pt](9.4,3.38)(14.35,3.4)
\psline[linewidth=.3pt](9.4,1.88)(14.35,1.9)
\psline{->}(0.4,0.1)(0.38,4.77)
\psline{->}(9.72,0.1)(9.7,4.77)
\psline[linewidth=.8pt]{->}(1.11,2.06)(2.14,3.5)
\psline[linewidth=.8pt]{->}(2.73,3.47)(4.06,4.18)
\psline[linewidth=.8pt]{->}(1.79,1.11)(2.52,2.12)
\psline[linewidth=.8pt]{->}(2.92,2.11)(4.12,2.78)
\parametricplot[linewidth=.3pt]{1.7235674377538832}{2.718741414308379}{1*5.72*cos(t)+0*5.72*sin(t)+9.88|0*5.72*cos(t)+1*5.72*sin(t)+3.18}
\parametricplot[linewidth=.3pt]{1.472907948870405}{2.949113549971398}{1*3.37*cos(t)+0*3.37*sin(t)+10.26|0*3.37*cos(t)+1*3.37*sin(t)+4.11}
\psline[linewidth=.8pt]{->}(4.95,6.09)(5.8,7.53)
\psline[linewidth=.8pt]{->}(6.29,7.63)(7.52,8.62)
\psline[linewidth=.8pt]{->}(7.11,5.31)(7.62,6.65)
\psline[linewidth=.8pt]{->}(8.34,6.88)(9.53,7.71)
\psline[linewidth=.8pt]{->}(9.85,1.88)(11.34,1.89)
\psline[linewidth=.8pt]{->}(12.35,1.89)(13.84,1.9)
\psline[linewidth=.8pt]{->}(10.04,3.38)(11.53,3.39)
\psline[linewidth=.8pt]{->}(12.42,3.39)(13.91,3.4)
\parametricplot{1.7448508960176878}{2.9743639964300783}{1*3.26*cos(t)+0*3.26*sin(t)+5.02|0*3.26*cos(t)+1*3.26*sin(t)+4.47}
\psline{->}(1.81,5.01)(1.79,4.91)
\parametricplot{-0.36142931821872804}{1.293687305190943}{1*2.19*cos(t)+0*2.19*sin(t)+10.49|0*2.19*cos(t)+1*2.19*sin(t)+5.42}
\psline{->}(12.54,4.64)(12.5,4.53)
\rput[bl](1.1,6.5){$\{q^\mu\}$}
\rput[bl](12.6,6.5){$\{Q^\mu\}$}
\rput[bl](7.5,7.2){$\boldsymbol\xi$}
\rput[bl](7.,4.1){$U$}
\rput[bl](2,2.3){$(\xi^\mu)$}
\rput[bl](11.5,2.4){$(\delta^{\mu\alpha})$}
\psdots[dotstyle=*](7.92,6.54)
\rput[bl](8.05,6){$p$}
\psdots[dotstyle=*](9.67,7.43)
\rput[bl](9.74,6.8){$p_\varepsilon$}
\psdots[dotstyle=*](2.7,1.98)
\rput[bl](2.76,1.4){$q^\mu$}
\psdots[dotstyle=*](4.12,2.52)
\rput[bl](4.15,1.9){$q^\mu_\varepsilon$}
\psdots[dotstyle=*](11.77,1.89)
\uput{.2}[-90](11.77,1.89){$Q^\mu$}
\psdots[dotstyle=*](14.04,1.9)
\uput{.2}[-90](14.04,1.9){$Q^\mu_\varepsilon$}
\end{pspicture*}
\caption{The vector field $\boldsymbol\xi$ over the domain $U$ and its representatives with respect to two extended coordinate systems $\{q^\mu\}$ and $\{Q^\mu\}$. The former is arbitrary while the latter rectifies $\boldsymbol\xi$ into an uniform field along one coordinate, namely $Q^\alpha$.}\label{fig:transformation}
\end{figure}
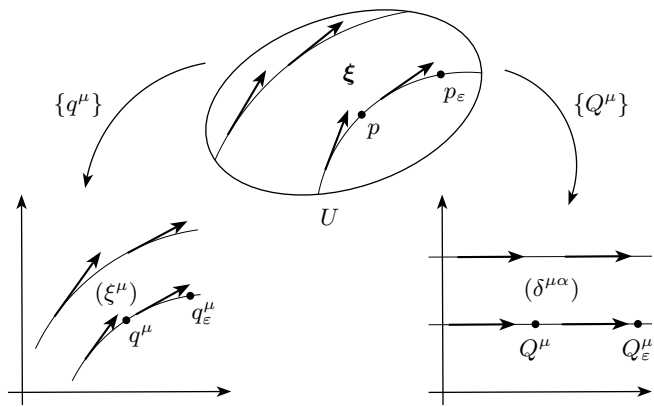

A \emph{continuous point transformation}\footnote{`Continuous transformation' is an accepted terminology in physics which suffers a lack of precision. Indeed, much more than continuity is involved in what follows.} $\Phi$ of $\mathcal E$ is, roughly speaking, a process which locally displaces its points in the flow of a smooth vector field $\boldsymbol\xi$, the \emph{generator} of $\Phi$ (see figure~\ref{fig:transformation}). To be a little more precise \cite{Chern}, any point $p$ is driven by $\Phi$ along a piece of integral curve $\varepsilon\mapsto p_\varepsilon$ of $\boldsymbol\xi$, the parameter $\varepsilon$ taking its values in some interval around zero, in such a way that the dependence in $\varepsilon$ is smooth and $p_0=p$. Actually, without any other precision, the interval has \textit{a priori} only a local character: all we can say is that there exists, everywhere, a connected open neighbourhood $U$ and an interval $(-\eta,\eta)$ such that all the points of $U$ follow, under the action of $\Phi$, a piece of an integral curve of $\boldsymbol\xi$ smoothly parametrised by $\varepsilon\in(-\eta,\eta)$. For this reason, we will work locally and focus ourselves on such a subset $U$ of the extended configuration space. Within this restriction, $\Phi$ is locally a smooth map from $U\times(-\eta,\eta)$ to $\mathcal E$ defined by $\Phi(p,\varepsilon)=p_\varepsilon$. It verifies the two following properties:
\begin{itemize}
\item $\Phi(p,0)=p$,
\item $\Phi(\Phi(p,\varepsilon),\varepsilon')=\Phi(p,\varepsilon+\varepsilon')$, 
\end{itemize}
whenever these expressions make sense.\footnote{Beyond the belonging of $p$ in $U$ and $\varepsilon,\varepsilon'$ in $(-\eta,\eta)$, one must be sure that $\Phi(p,\varepsilon)$ and $\varepsilon+\varepsilon'$ remain in $U$ and $(-\eta,\eta)$ respectively.}

For each fixed value of $\varepsilon$, the continuous transformation $\Phi$ induces a \emph{local diffeomorphism}\footnote{A diffeomorphism is a one-to-one mapping which is smooth as well as its inverse. By ̀`local diffeomorphism' in the present context is meant the existence, for any point $p\in U$, of an open neighbourhood $V$ of $p$ in $U$ such that each $\Phi_\varepsilon$ realizes a diffeomorphism from $V$ to some open subset of $\mathcal E$.} $\Phi_\varepsilon$ associating with each point $p$ of $U$ the point $\Phi(p,\varepsilon)$. When $\Phi$ can be conceived\footnote{Vector fields for which this is possible are said to be \emph{complete}.} as a map from $\mathcal E\times\mathbb R$ to $\mathcal E$ as wholes, it is clear that the collection $\{\Phi_\varepsilon\}$ forms a group of diffeomorphisms of $\mathcal E$ for the law $\Phi_\varepsilon\circ\Phi_{\varepsilon'}=\Phi_{\varepsilon+\varepsilon'}$, with $\Phi_0$ as identity, the inverse of each $\Phi_{\varepsilon}$ being $\Phi_{-\varepsilon}$. Otherwise, $\{\Phi_\varepsilon\}$ is said to be a \emph{local one-parameter group of local diffeomorphisms}.

\subsection{Transformations of evolutions}

Truncating $U$ if necessary, we will assume that it is the domain of an extended coordinate system $\{t,q^i\}$. It allows us to identify the points of $U$ with their coordinates in $\{t,q^i\}$. In some instances, we will denote $t$ by $q^0$ and adopt the convention that Greek indices cover the range between 0 and $n$. Under the action of $\Phi$, a given point $p$ of $U$ is `set in motion' whereas remaining in $U$ for sufficiently small values of $\lvert\varepsilon\rvert$. In coordinates one has thus\footnote{By definition, a function $f$ of $\varepsilon$ is a `little-o' of $\varepsilon$ if the ratio $f(\varepsilon)/\varepsilon$ tends to zero with $\varepsilon$.}
\begin{align*}
\Phi\colon\quad q^\mu\longmapsto q^\mu_\varepsilon=q^\mu+\varepsilon\,\xi^\mu(t,q^i)+\text{o}(\varepsilon),
\end{align*} 
where $(\xi^\mu)=(\tau,\xi^i)$ are the components of $\boldsymbol\xi$ in the system $\{q^\mu\}=\{t,q^i\}$.

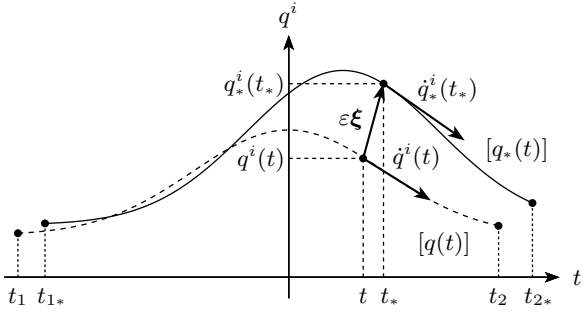
\begin{figure}
\begin{center}
\psset{xunit=1.8cm,yunit=1.4cm,algebraic=true,dimen=middle,dotstyle=o,dotsize=3pt 0,linewidth=0.5pt,arrowsize=3pt 2,arrowinset=0.25}
\begin{pspicture*}(-2.1,-0.6)(2.2,2.4)
\psaxes[labelFontSize=\scriptstyle,xAxis=true,yAxis=true,labels=none,ticks=none]{->}(-0.0,-.3)(-2.1,-0.5)(2.,2.)
\psplot[plotpoints=200,linestyle=dashed,dash=2pt 2pt]{-2}{1.55}{2.718281828459045^(-x^(2.0))+0.1}
\psplot[plotpoints=200]{-1.8}{1.8}{1.2*2.718281828459045^(-(x-0.4)^(2.0)+0.2)+0.2}
\psdots[dotstyle=*](0.55,0.83)
\psline[linestyle=dashed,dash= 1.5pt 1.5pt](0.55,0.83)(.55,-.3)
\uput{0.15cm}[-90](.55,-.3){$t$}
\psline[linestyle=dashed,dash= 1.5pt 1.5pt](0.55,0.83)(0,0.83)
\uput{0.05cm}[180](0,0.83){{$q^i(t)$}}
\psline[linewidth=.8pt]{->}(.55,.83)(1.05,.43)
\uput{0.4cm}[105](1.05,.43){{$\dot q^i(t)$}}
\psdots[dotstyle=*](0.7,1.54)
\psline[linewidth=.8pt]{->}(0.55,0.83)(0.7,1.54)
\psline[linestyle=dashed,dash= 1.5pt 1.5pt](0.7,1.54)(-.0,1.54)
\uput{.05cm}[180](-.0,1.54){$q^i_\sast(t_\sast)$}
\psline[linestyle=dashed,dash= 1.5pt 1.5pt](0.7,1.54)(0.7,-.3)
\uput{.15cm}[-90](0.75,-.3){$t_{\sast}$}
\psline[linewidth=.8pt]{->}(0.7,1.54)(1.3,1.0)
\rput[bl](.95,1.37){$\dot q^i_\sast(t_\sast)$}
\rput[bl](.95,-.1){{$[q(t)]$}}
\rput[bl](1.45,.8){{$[q_\sast(t)]$}}
\rput[bl](.37,1.1){$\varepsilon\boldsymbol\xi$}
\uput[0](2,-.3){$t$}
\uput{.1cm}[90](0.0,2){$q^i$}
\psdots[dotstyle=*](-2,0.1183)
\psline[linestyle=dashed,dash= 1pt 1pt](-2,0.1183)(-2,-.3)
\uput{.15cm}[-90](-2.0,-.3){$t_1$}
\psdots[dotstyle=*](-1.8,0.212)
\psline[linestyle=dashed,dash= 1pt 1pt](-1.8,0.212)(-1.8,-.3)
\uput{.15cm}[-90](-1.75,-.3){$t_{1\sast}$}
\psdots[dotstyle=*](1.8,0.406)
\psline[linestyle=dashed,dash= 1pt 1pt](1.8,0.406)(1.8,-.3)
\uput{.15cm}[-90](1.85,-.3){$t_{2\sast}$}
\psdots[dotstyle=*](1.55,0.19)
\psline[linestyle=dashed,dash= 1pt 1pt](1.55,0.19)(1.55,-.3)
\uput{.15cm}[-90](1.55,-.3){$t_{2}$}
\end{pspicture*}
\end{center} 
\caption{Under the action of $\Phi$, and for a sufficiently small value of $\lvert\varepsilon\rvert$, the original evolution $[q(t)]$ (dashed line) is mapped to another one, $[q_\sast(t)]$ (solid line). Here, the infinitesimal limit is sketched by exagerating the distance between the two evolutions.}\label{fig:transformation_evolution}
\end{figure}

In what follows, we will be concerned by the effect of $\Phi$ on a local smooth evolution in $\mathcal M$ between two extremities of time $t_1$ and $t_2$. It naturally traces a graph $t\mapsto p(t)=(t,q(t))$ in $\mathcal E$ which will be assumed contained in $U$. Let us denote it by $[q(t)]$. Provided that $\lvert\varepsilon\rvert$ is sufficiently small, the graph is transformed to some curve lying again in $U$. Shrinking once more the interval of $\varepsilon$ values around zero, if necessary, the transformed curve is the graph of another evolution.\footnote{To be more precise, the function $t\mapsto \tau(t,q^i(t))$ is smooth because $\tau$ and $t\mapsto q(t)$ are so. Its continuous derivative $\dot\tau$ is thus bounded on $[t_1,t_2]$. Let $\alpha$ be an upper bound of $\lvert\dot\tau\rvert$ between $t_1$ and $t_2$. For a fixed value $\varepsilon$ such that $\lvert\varepsilon\rvert<\alpha^{-1}$, the transform $t_\varepsilon$ of $t$ verifies
\begin{align*}
\frac{\rm d t_\varepsilon}{\rmd t}=1+\varepsilon\,\dot\tau>1-\lvert\varepsilon\dot\tau\rvert>0
\end{align*}
during this range of time. Since $t_\varepsilon$ increases strictly with $t$, the transformed curve is the graph of some evolution.} From now on, $\varepsilon$ will be treated as an infinitesimal and its little-o will be systematically ignored. Consequently, the graph $t\mapsto p(t)$ is infinitesimally transformed by $\Phi$ into the graph $t\mapsto p_\sast(t)=(t,q_\sast(t))$ of another evolution $[q_\sast(t)]$ according to
\begin{align*}
\Phi\colon\quad p(t)=(t,q(t))\longmapsto p_\sast(t_\sast)=(t_\sast,q_\sast(t_\sast)).
\end{align*}
In coordinates, one has thus
\begin{align*}
(t_{\sast},q^i_{\sast}(t_{\sast}))=(t+\varepsilon\,\tau(t,q^i(t)),q^i(t)+\varepsilon\,\xi^i(t,q^j(t))).
\end{align*}
The transformed evolution $[q_\sast(t)]$ takes place between the two extremities of time
\begin{align*}
t_{1\sast}\coloneqq t_1+\varepsilon\,\tau(t_1,q^i(t_1))\;\;\;\text{and}\;\;\; t_{2\sast}\coloneqq t_2+\varepsilon\,\tau(t_2,q^i(t_2)).
\end{align*}
The transformation is illustrated in figure~\eqref{fig:transformation_evolution}. 

As usual, we will symbolise the total $t$-derivative by an overdot. Since $t_{\sast}$ and $q_{\sast}(t_{\sast})$, seen as implicit functions of $t$ along $[q(t)]$, are obviously differentiable, the velocities of $[q_{\sast}(t)]$ at the instant $t_{\sast}$ are well-defined by the chain rule:
\begin{align*}
\dot{q}^i_{\sast}(t_{\sast})=\frac{\rmd}{\rmd t}\big(q^i_{\sast}(t_{\sast})\big)\Big\slash\frac{\rmd t_\sast}{\rmd t_{\phantom\sast}}=\dot q^i(t)+\varepsilon\big(\dot\xi^i-\dot q^i(t)\dot\tau\big),
\end{align*}
where the quantities in the right-hand side are tacitly evaluated at the instant $t$ along $[q(t)]$. It is clear that the transform $[q_{\sast}(t)]$ is smooth as well. Then, under the further assumption that $[q(t)]$ is $\mathscr C^2$, the second derivatives of $[q_{\sast}(t)]$ exist at the instant $t_\sast$ and are, by the same rule:
\begin{align*}
\ddot{q}^i_{\sast}(t_\sast)=\ddot q^i(t)+\varepsilon(\ddot\xi^i-\dot q^i(t)\ddot\tau-2\ddot q^i(t)\dot\tau\big),
\end{align*}
with the same tacit evaluation. The process can be iterated as many time as $[q(t)]$ is differentiable, and if $[q(t)]$ is $\mathscr C^k$ so is its transform. For notational convenience when evolutions are considered, we will assume that when $q$, $q^i$, $\dot q^i$, etc. (resp. $q_\sast$, $q^i_\sast$, $\dot q^i_\sast$, etc.) come without argument, they are evaluated at the instant $t$ (resp. $t_\sast$).

Before going further, let us say a few words on extended coordinate systems. In the system $\{t,q^i\}$ used up to now there is naturally an asymmetry between the time $t$ and the coordinates $q^i$: the former is the independent variable whereas the latter  are the dependent ones, in the sense that one is interested by evolutions of the $q^i$ as functions of $t$. However, nothing prevents us from performing a change of extended coordinate system $\{q^\mu\}\to\{q^{\mu'}\}$. If we suppose that, along the evolutions under consideration, $q^{0'}$ strictly increases with $t$ then it can be taken as a new independent variable, let us say a new time $t'$. The above analysis could have been done by means of the system $\{t',q^{i'}\}$, without any change in its form: all we have to do is to `put primes on the indices' and to consider the total $t'$-derivative (which we can associate with another symbol than the dot if we wonder about possible confusions). In particular, this implies the contravariant transformation of $\boldsymbol\xi$'s components:
\begin{align*}
\xi^\mu\longrightarrow \xi^{\mu'}=\frac{\partial q^{\mu'}}{\partial q^\nu}\,\xi^\nu,
\end{align*}
where the Einstein summation convention is assumed (as it will be throughout the present article).

\subsection{Prolongations and symmetries}

The generator $\boldsymbol\xi$ has a natural action on scalar fields by evaluating their rate of change in its direction. Let $\mathscr G_0(t,q)$ be a scalar field in $U$. It is an absolute object represented in the system $\{t,q^i\}$ by the function $G_0(t,q^i)$ such that $G_0(t,q^i)=\mathscr G_0(t,q)$. Whereas $(t,q)$ is infinitesimally transformed into $(t_\sast,q_\sast)$, the value taken by $\mathscr G_0$ undergoes the variation
\begin{align*}
\delta_{\boldsymbol\xi}\mathscr G_0(t,q)=\mathscr G_0(t_\sast,q_\sast)-\mathscr G_0(t,q)=\varepsilon\,\boldsymbol\xi(\mathscr G_0)(t,q).
\end{align*}
One has thus in coordinates
\begin{align*}
\delta_{\boldsymbol\xi}\mathscr G_0(t,q)=G_0(t_\sast,q^i_\sast)-G_0(t,q^i)=\varepsilon\,\xi^\mu\frac{\partial G_0}{\partial q^\mu}(t,q^i).
\end{align*}
The fact that $\mathscr G_0$ and $G_0$ are locally `the same thing' allows us to identify as usual $\boldsymbol\xi$ with the operator $\xi^\mu\partial_\mu$ and to write indifferently
\begin{align*}
\delta_{\boldsymbol\xi}\mathscr G_0=\varepsilon\,\boldsymbol\xi(\mathscr G_0)=\varepsilon\,\boldsymbol\xi(G_0)=\delta_{\boldsymbol\xi}(G_0)\quad,\quad\boldsymbol\xi=\xi^\mu\partial_\mu\,.
\end{align*}
The use of an extended coordinate system provides an expression to $\boldsymbol\xi$ whose form does not depend on the system used. Indeed, identifying $\mathscr G_0$ with its representative in a primed system yields obviously
\begin{align*}
\boldsymbol\xi=\xi^\mu\partial_\mu=\xi^{\mu'}\partial_{\mu'}\,.
\end{align*}

Now, consider some absolute scalar quantity $\mathscr G_1(t,q,\dot q)$ which depends also on the velocities (the precise space in which that quantity lives will not be our concern). Let $G_1(t,q^i,\dot q^i)$ be its representative by means of $\{t,q^i\}$. Whereas the evolution $[q(t)]$ is infinitesimally transformed into $[q_{\sast}(t)]$, the value taken by $\mathscr G_1$ undergoes, when passing from the point $(t,q(t))$ to its image, the variation
\begin{align*}
\delta_{\boldsymbol\xi}\mathscr G_1(t,q,\dot q)=G_1(t_{\sast},q_{\sast},\dot q^i_\sast)- G_1(t,q^i,\dot q^i)
\end{align*}
which leads to
\begin{align*}
\delta_{\boldsymbol\xi}\mathscr G_1=\varepsilon\,\boldsymbol \xi^{[1]}(\mathscr G_1)=\varepsilon\,\boldsymbol \xi^{[1]}(G_1)=\delta_{\boldsymbol\xi}G_1,
\end{align*}
where
\begin{align}
\boldsymbol \xi^{[1]}=\boldsymbol\xi+(\dot \xi^i-\dot q^i\dot\tau)\frac{\partial}{\partial\dot q^i}\label{first_prolongation}
\end{align}
is the so-called \emph{first prolongation} of $\boldsymbol \xi$. For the same reason as before, this operator is an absolute quantity whose expression is form invariant: by means of a primed system, one has again
\begin{align*}
\boldsymbol \xi^{[1]}=\boldsymbol\xi+(\mathring \xi^{i'}-\mathring q^{i'}\mathring\tau')\frac{\partial}{\partial\mathring q^{i'}},
\end{align*}
where the $t'$-derivative has been symbolised by an empty bullet. It is quite easy to verify that the prolongation is compatible with the Lie algebra of vector fields in the sense that
\begin{align}
\big(\boldsymbol\xi_1+\lambda\boldsymbol\xi_2)^{[1]}&=\boldsymbol\xi_1^{[1]}+\lambda\boldsymbol\xi_2^{[1]},\label{compatibility_ev}\\
\big[\boldsymbol\xi_1,\boldsymbol\xi_2\big]^{[1]}&=\big[\boldsymbol\xi_1^{[1]},\boldsymbol\xi_2^{[1]}\big]\label{compatibility_bracket},
\end{align}
where $\boldsymbol\xi_1$ and $\boldsymbol\xi_2$ are two vector fields, $\lambda$ is a constant, and where the bracket has always the usual meaning of a commutator. For scalar quantities $\mathscr G_2(t,q,\dot q,\ddot q)$, the variation will be evaluated by the \emph{second prolongation} 
\begin{align*}
\boldsymbol\xi^{[2]}\coloneqq\boldsymbol\xi^{[1]}+(\ddot\xi^i-\dot q^i\ddot\tau-2\ddot q^i\dot\tau)\frac{\partial}{\partial\ddot q^{\,i}}\,,
\end{align*}
and so on. Successive prolongations $\boldsymbol \xi^{[k]}$ may be recursively defined to evaluate the variation of any scalar quantity $\mathscr G_k$ of $t$, $q$, $\dot q$, $\ddot q$, and so forth, up to the derivatives of order $k$ (provided that the considered evolutions are sufficiently smooth). This integer $k$ will be called the order of $\mathscr G_k$ (with the convention that an order zero corresponds to functions of $t$ and $q$). The expressions of the prolongations are by construction form invariants and it can be recursively verified that they remain compatible with the Lie algebra to all orders.\footnote{This property might however be established directly in an intrinsic way but it would necessitate a deeper knowledge of the underlying geometry \cite{Olver}.}

One says that the transformation $\Phi$ is a symmetry of a scalar quantity $\mathscr G$ if it leaves its value invariant up to the first order in $\varepsilon$ for any sufficiently smooth evolution. If $\mathscr G$ is of order $k$, the symmetry condition is thus
\begin{align}
\boldsymbol\xi^{[k]}(\mathscr G)=0,\label{symmetry}
\end{align}
seen as an identity in $t$, $q$, and all the involved derivatives of $q$ (with the convention $\boldsymbol\xi^{[0]}\coloneqq\boldsymbol\xi$). Therefore, it imposes a restriction on the algebraic expression of $\mathscr G$. Such a constraint is an expected feature of the concept of symmetry. Furthermore, by the compatibility between the prolongations and the Lie algebra, it is clear that the generators of the symmetries form by themselves a Lie algebra (of the symmetry group).

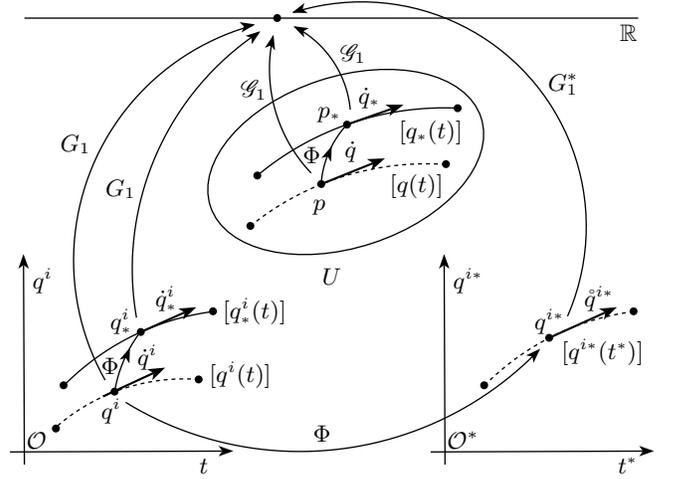
\begin{figure}
\centering
\psset{xunit=.6cm,yunit=.6cm,algebraic=true,dotstyle=o,dotsize=3pt 0,linewidth=0.5pt,arrowsize=3pt 2,arrowinset=0.25}
\begin{pspicture*}(0.05,-0.15)(14.4,10.4)
\psline{->}(0.08,0.38)(5.03,0.4)
\psline{->}(9.4,0.38)(14.35,0.4)
\rput{18.19}(7.52,6.81){\psellipse(0,0)(3.17,1.91)}
\psline{->}(0.4,0.1)(0.38,4.77)
\psline{->}(9.72,0.1)(9.7,4.77)
\parametricplot[linestyle=dashed,dash=1.5pt 1.5pt]{1.5023735446043573}{2.250976851795139}{1*6.21*cos(t)+0*6.21*sin(t)+9.31|0*6.21*cos(t)+1*6.21*sin(t)+0.56}
\parametricplot{1.5242092476289315}{2.266682184671147}{1*6.46*cos(t)+0*6.46*sin(t)+9.7|0*6.46*cos(t)+1*6.46*sin(t)+1.55}
\parametricplot{2.390413266560704}{3.071547148424186}{1*2.16*cos(t)+0*2.16*sin(t)+9.13|0*2.16*cos(t)+1*2.16*sin(t)+6.17}
\psline[linewidth=.8pt]{->}(7.55,7.64)(8.8,8.08)
\psline[linewidth=.8pt]{->}(6.98,6.32)(8.39,6.89)
\parametricplot[linestyle=dashed,dash=1.5pt 1.5pt]{1.4951704260121637}{2.28}{1*4.3*cos(t)+0*4.3*sin(t)+3.94|0*4.3*cos(t)+1*4.3*sin(t)+-2.3}
\parametricplot{1.7003723203199148}{2.36277354312733}{1*5.67*cos(t)+0*5.67*sin(t)+5.31|0*5.67*cos(t)+1*5.67*sin(t)+-2.13}
\parametricplot[linestyle=dashed,dash=1.5pt 1.5pt]{1.7003723203199153}{2.3627735431273305}{1*5.67*cos(t)+0*5.67*sin(t)+14.63|0*5.67*cos(t)+1*5.67*sin(t)+-2.13}
\psline[linewidth=.8pt]{->}(12.03,2.91)(13.53,3.59)
\psline[linewidth=.8pt]{->}(2.14,1.61)(3.52,2.25)
\psline[linewidth=.8pt]{->}(2.98,3.04)(4.21,3.59)
\parametricplot{4.161815214463326}{5.503153999621527}{1*7.37*cos(t)+0*7.37*sin(t)+6.5|0*7.37*cos(t)+1*7.37*sin(t)+7.76}
\psline(0.4,10)(14,10)
\parametricplot{0.09066658413989785}{1.0870262104624888}{1*2.31*cos(t)+0*2.31*sin(t)+5.32|0*2.31*cos(t)+1*2.31*sin(t)+7.76}
\parametricplot{2.9235074130954937}{3.9030973737663173}{1*3.34*cos(t)+0*3.34*sin(t)+9.17|0*3.34*cos(t)+1*3.34*sin(t)+8.87}
\parametricplot{2.159290243388058}{3.313925237400893}{1*6.31*cos(t)+0*6.31*sin(t)+9.1|0*6.31*cos(t)+1*6.31*sin(t)+4.44}
\parametricplot{1.8372431721317548}{3.6561195736011443}{1*5.4*cos(t)+0*5.4*sin(t)+6.89|0*5.4*cos(t)+1*5.4*sin(t)+4.63}
\parametricplot{-0.41118318948452703}{1.8780964374124458}{1*4.98*cos(t)+0*4.98*sin(t)+7.87|0*4.98*cos(t)+1*4.98*sin(t)+5.38}
\psline{->}(5.59,9.68)(5.63,9.71)
\psline{->}(5.46,9.85)(5.51,9.86)
\psline{->}(5.91,9.59)(5.92,9.63)
\psline{->}(6.4,9.8)(6.36,9.82)
\psline{->}(6.36,10.13)(6.33,10.12)
\psline{->}(7.17,7.06)(7.21,7.16)
\parametricplot{2.4871835782169716}{2.975389829374362}{1*2.99*cos(t)+0*2.99*sin(t)+5.35|0*2.99*cos(t)+1*2.99*sin(t)+1.23}
\psline{->}(11.74,2.58)(11.83,2.67)
\psline{->}(2.72,2.64)(2.74,2.69)
\rput[bl](13.6,9.5){$\mathbb R$}
\rput[bl](7.,4.1){$U$}
\rput[bl](4.28,-.1){$t$}
\rput[bl](13.55,-.1){$t^\sast$}
\rput[bl](.58,3.9){$q^i$}
\rput[bl](9.9,3.9){$q^{i\sast}$}
\psdots[dotstyle=*](5.41,5.39)
\psdots[dotstyle=*](9.74,6.76)
\psdots[dotstyle=*](5.56,6.51)
\psdots[dotstyle=*](10,8)
\psdots[dotstyle=*](7.55,7.64)
\rput[bl](6.9,7.7){$p_\sast$}
\psdots[dotstyle=*](6.98,6.32)
\rput[bl](6.8,5.65){$p$}
\rput[bl](5.2,8.2){$\mathscr G_1$}
\rput[bl](7.4,9){$\mathscr G_1$}
\rput[bl](7.5,6.8){$\dot q$}
\rput[bl](7.8,8.){$\dot q_\sast$}
\rput[bl](6.57,6.78){$\Phi$}
\rput[bl](6.8,.6){$\Phi$}
\psdots[dotstyle=*](1.1,0.9)
\psdots[dotstyle=*](2.4,1.72)
\rput[bl](2.1,1.){$q^i$}
\rput[bl](2.9,2.2){$\dot q^i$}
\rput[bl](1.2,7){$G_1$}
\rput[bl](2.2,6){$G_1$}
\rput[bl](12,8.3){$G_1^*$}
\rput[bl](8.5,6){$[q(t)]$}
\rput[bl](8.7,7.2){$[q_\sast(t)]$}
\rput[bl](4.5,1.8){$[q^i(t)]$}
\rput[bl](4.8,3.2){$[q_\sast^i(t)]$}
\rput[bl](12.3,2.3){$[q^{i\sast}(t^\sast)]$}
\psdots[dotstyle=*](4.26,1.99)
\psdots[dotstyle=*](1.27,1.86)
\psdots[dotstyle=*](4.58,3.5)
\psdots[dotstyle=*](2.98,3.04)
\rput[bl](2.3,3){$q_\sast^i$}
\rput[bl](3.3,3.5){$\dot q_\sast^i$}
\psdots[dotstyle=*](10.59,1.86)
\psdots[dotstyle=*](12.03,2.91)
\psdots[dotstyle=*](13.9,3.5)
\rput[bl](11.7,3){$q^{i\sast}$}
\rput[bl](12.8,3.5){$\mathring q^{i\sast}$}
\psdots[dotstyle=*](6,10)
\rput[bl](2.1,2.1){$\Phi$}
\rput[bl](0.45,0.45){$\mathcal O$}
\rput[bl](9.77,.45){$\mathcal O^*$}
\end{pspicture*}
\caption{In terms of representations in coordinates, the active viewpoint of $\Phi$ is the transcription of its action in an arbitrary extended system $\{t,q^i\}$. As for the passive viewpoint, it converts $\Phi$ into a change of extended system~\eqref{passive_transform} such that the evolution $[q(t)]$ is seen by the new observer $\mathcal O^*$ as it would appear transformed by $\Phi$ to the original observer $\mathcal O$. The absolute quantity $\mathscr G_1$, depending on the event and the velocity, admits $\Phi$ as symmetry: its values are left invariant by the transformation. One can read the active viewpoint as the expression of this property in an arbitrary system through a representative $G_1$. The passive viewpoint is the functional invariance of the representative under the change of extended system generated by $\Phi$.}\label{fig:viewpoints}
\end{figure}

\subsection{The meaning of a point symmetry in the passive viewpoint}

Up to now, the mapping $q^\mu\to q^\mu+\varepsilon\xi^\mu$ was envisaged as a representation of the infinitesimal action of $\Phi$ in a certain extended coordinate system $\{q^\mu\}=\{t,q^i\}$. This is an active transformation. The passive counterpart is reached when this mapping is rather seen as a change of extended coordinates (also generated by $\Phi$). To avoid any confusion between the objects in question, we will denote by $\{q^{\mu\sast}\}=\{t^\sast,q^{i\sast}\}$ the new extended system, thereby related to $\{q^\mu\}$ through
\begin{align}
q^{\mu\sast}=q^\mu+\varepsilon\,\xi^\mu.\label{passive_transform}
\end{align}

An observer $\mathcal O^*$, equipped with the system $\{q^{\mu\sast}\}$, associates with each point $p$ the coordinates that an observer ${\mathcal O}$, equipped with $\{q^{\mu}\}$, would have associated with its transform $p_\sast$ (see figure~\ref{fig:viewpoints}).
Put it differently, everything appears for $\mathcal O^*$ as if it were transformed by $\Phi$ from the point of view of $\mathcal O$. While the latter describes an evolution $[q(t)]$ by a graph $t\mapsto (t,q^i(t))$ between $t_1$ and $t_2$ which will be denoted by $[q^i(t)]$, the former traces, between $t^\sast_1=t_{1\sast}$ and $t^\sast_2=t_{2\sast}$, the graph $t^\sast\mapsto(t^\sast,q^{i\sast}(t^\sast))=(t_{\sast},q^i_{\sast}(t_{\sast}))$ that we will denote by $[q^{i\sast}(t^{\sast})]$. In particular, the derivatives of $[q^{i\sast}(t^\sast)]$ as seen from $\mathcal O^*$ at the instant $t^{\sast}$ coincide with the ones of $[q^i_\sast(t)]$ as seen from $\mathcal O$ at the instant $t_{\sast}$:
\begin{align*}
\mathring{q}^{i\sast}(t^{\sast})=\dot q^i_{\sast}(t_\sast)\quad,\quad\ringring{q}^{\,i\sast}(t^\sast)=\ddot q^{\,i}_\sast(t_\sast)\quad,\quad\text{etc.}
\end{align*}
where the total $t^\sast$-derivative has been symbolised by an empty bullet. Once again, for notational convenience when evolutions are considered, we will assume that when $q^{i\sast}$, $\mathring q^{i\sast}$, etc. come without argument, they are evaluated at the instant $t^\sast$, just like $q^i$, $\dot q^i$, etc. are at $t$ and $q^i_\sast$, $\dot q^i_\sast$, etc. at $t_\sast$.

Now, some absolute quantity $\mathscr G$ represented by a function $G$ in the system $\{q^\mu\}$ will be represented by $G^*$ in $\{q^{\mu\sast}\}$ according to the correspondence
\begin{align}
G^*(t^\sast,q^{i\sast},\mathring q^{i\sast},\dots)&=G(t,q^i,\dot q^i,\dots)\label{G_Gtransf}\\
&=G(t_\sast,q^i_\sast,\dot q^i_\sast,\dots)-\delta_{\boldsymbol\xi}G(t,q^i,\dot q^i,\dots)\notag.
\end{align}
Hence, the symmetry criterion~\eqref{symmetry} amounts to the equality
\begin{align}
G^*(t^\sast,q^{i\sast},\mathring q^{i\sast},\dots)=G(t_\sast,q^i_\sast,\dot q^i_\sast,\dots).\label{equality_G}
\end{align}
Since, in both sides, the arguments are numerically equal, one concludes that $G^*$ must be the same function of $(t,q^i,\dot q^i,\dots)$ than $G$ is of $(t^\sast,q^{i\sast},\mathring q^{i\sast},\dots)$. In the passive point of view, the symmetry manifests itself as a functional invariance under the change of coordinates generated by $\Phi$: the two observers $\mathcal O$ and $\mathcal O^*$ both describe $\mathscr G$ in the same terms. We emphasize that the equality~\eqref{equality_G} will not be condensed into $G^*=G$ since the latter commonly designates the `transformation law (of representatives) of scalars~\eqref{G_Gtransf} in the tradition of physics.\footnote{Physicists attach importance to the variables because they are supposed to be charged of meaning. In particular, when they introduce a function, they nearly always think about a `function of'. This \textit{habitus} is often convenient but can sometimes lead to misconceptions as well as needless complications. In the present paper, we have decided for pedagogical reasons to not depart from this \textit{habitus}. This is why we insist upon the `functional equality' of two `functions of' instead of simply the `equality' of two `functions' as mathematicians would say.}

\subsection{Point symmetries and adapted coordinates}

A well-known theorem of differential geometry \cite{Chern} states that, around a point of $U$ where $\boldsymbol\xi$ does not vanish, one can always find a peculiar coordinate system $\{Q^\mu\}$ for which $\boldsymbol\xi$ reduces to the partial derivative with respect to one of the $Q^\mu$, say $Q^\alpha$. Geometrically speaking, it means that this system --- which is said to be \emph{adapted} to $\boldsymbol\xi$ (or to the transformation $\Phi$) --- is such that the $Q^\alpha$ coordinate lines coincide locally with the integral curves of $\boldsymbol\xi$ (see figure~\ref{fig:transformation}). It has the advantage of transcribing the transformation $\Phi$ as a mere translation of magnitude $\varepsilon$ in the direction of $Q^\alpha$ (up to the first order in $\varepsilon$):
\begin{align*}
Q^\mu\longrightarrow Q^\mu+\varepsilon\,\delta^{\mu\alpha}.
\end{align*}

In the case where $\dot Q^0$ is strictly increasing along the evolutions under consideration, it can be used as a new time $T$. The resulting system $\{T,Q^i\}$ trivializes the prolongations of $\boldsymbol\xi$: all of them reduce to the partial derivative with respect to $Q^\alpha$, too. The symmetry condition~\eqref{symmetry} is now synonym for an independence on $Q^\alpha$ in any case, that is, a translational invariance along the direction of $Q^\alpha$. If $\alpha=0$, the symmetry means an explicit independence on the new time. Otherwise, if $\alpha>0$, it means an independence on a coordinate $Q^i$, although it does not prevent at all from a dependence on its derivatives with respect to $T$. In particular, if we are in position to permute $Q^0$ with a coordinate $Q^i$, one can transform a time independence into a coordinate one, and \textit{vice versa}.

\section{Noether point symmetries}\label{sec:Noether}

\subsection{The Lagrangian framework}

Suppose that the dynamics is entirely governed by a $\mathscr C^2$ Lagrangian $L$ in the system $\{t,q^i\}$. The motions are the evolutions leaving stationary the action functional
\begin{align*}
S[q(t)]\coloneqq\int L(t,q^i,\dot q^i)\,\rmd t
\end{align*}
under the well-known conditions of Hamilton's principle. It amounts to say that they are solutions of the Euler-Lagrange equations
\begin{align}
\mathsf E_i(L)=0\qquad(i=1,\dots,n)\label{EL}
\end{align}
where
\begin{align*}
\mathsf E_i\coloneqq\frac{\partial}{\partial q^i}-\frac{\rmd}{\rmd t}\frac{\partial}{\partial\dot q^i}
\end{align*}
is the Lagrangian operator (or variational derivative) associated with the coordinate $q^i$. Following Noether, the quantities $\mathsf E_i(L)$ will be called the \emph{Lagrangian expressions} (\emph{Lagrangeschen Ausdr\"ucke}) of the variational problem in the coordinates $\{t,q^i\}$. In her own words, they are the `left-hand side of the Lagrangian equations' (\textit{die linken Seiten der Lagrangeschen Gleichungen}).

We recall that the action is a functional of the evolutions in $\mathcal M$ and, as such, is independent of the choice of extended coordinate system used for concrete calculations. Therefore, under a coordinate transformation $\{t,q^i\}\to\{t',q^{i'}\}$, the Lagrangian undergoes
\begin{align}
L\longrightarrow L'=L\,\frac{\rmd t\phantom{'}}{\rmd t'}\label{coordinate_change_L}
\end{align}
and~\eqref{EL} is equivalent to the set of Euler-Lagrange equations $\mathsf E_{i'}(L')=0$, where $\mathsf E_{i'}$ is the Lagrange operator associated with $q^{i'}$ in the new system. This equivalence is made explicit by the identity
\begin{align}
\mathsf E_{i'}(L')=\bigg(\frac{\partial q^j}{\partial q^{i'}}-\dot q^j\,\frac{\partial t}{\partial q^{i'}}\bigg)\frac{\rmd t\phantom{'}}{\rmd t'}\,\mathsf E_j(L)\label{covariance}
\end{align}
between the Lagrangian expressions. In textbooks, the above formula is rarely presented in its full generality and often restricted to the case where the time is not transformed. It can be verified by an explicit computation but a simpler proof will be given below, in subsection~\ref{subsec:parametrization}.

In addition to that `extended covariance property', the formalism is also characterized by its invariance under Lagrangian gauge transformations
\begin{align}
L\longrightarrow \overline L= L+\frac{\rmd}{\rmd t}\Big[\Lambda(t,q^i)\Big]\label{gauge_transformation}
\end{align}
since they do not affect the Euler-Lagrange equations. More precisely, two Lagrangians $L$ and $\overline L$ generate exactly the same Lagrange expressions (in an arbitrary coordinate system), i.e. are such that one has identically
\begin{align*}
\mathsf E_i(L)=\mathsf E_i(\overline L),
\end{align*}
if and only if (iff) they are related by a gauge transformation~\eqref{gauge_transformation}. In this case, they are said to be equivalent (indeed, we are faced with an equivalence relation in the mathematical sense of the term). The reason for this lies in the fact that a function $G(t,q^i,\dot q^i)$ verifies the $n$ identities $\mathsf E_i(G)=0$ iff it is the total $t$-derivative of some function of $t$ and $q^i$ \cite{Jose_Saletan}. We emphasize that the aforementioned equivalence is much more stronger than simply the equivalence of the Euler-Lagrange equations whereby $\{\mathsf E_i(L)=0\}$ and $\{\mathsf E_i(\overline L)=0\}$ have the same solutions. The simplest case where this weaker condition is encountered is when $\overline L$ differs from $L$ by a multiplicative nonzero constant factor. This point will be central below.

\subsection{The definition of Noether point symmetries and their meanings}\label{subsec:Noether}

The transformation $\Phi$ is a \emph{Noether point symmetry} of the variational problem if there exists a scalar field $\mathscr F(t,q)$ verifying, up to the first order in $\varepsilon$,
\begin{align}
\delta_{\boldsymbol\xi}S[q(t)]\coloneqq S[q_{\sast}(t)]-S[q(t)]=\varepsilon\Big[\mathscr F\Big]_{t_1}^{t_2}\label{variation}
\end{align}
for any evolution $[q(t)]$ \cite{Logan}. In Noether's original paper, the action was `only' assumed invariant in value: $\delta_{\boldsymbol\xi}S=0$. In this special case, the symmetry is said to be \emph{strict}. Otherwise, the field $\mathscr F$ will be called its \emph{Bessel-Hagen term} since the possibility of such symmetries was first envisaged in \textcite{Bessel-Hagen}. Obviously, $\mathscr F$ is given up to a meaningless constant and strict Noether point symmetries are those which admit a zero Bessel-Hagen term.

Suppose that $\Phi$ is a Noether point symmetry (NPS) with Bessel-Hagen (BH) term $\mathscr F$ and consider an evolution $[q(t)]$. Any variation $[\delta q_{\sast}(t)]$ of $[q_{\sast}(t)]$ keeping fixed its endpoints is, by the inverse transformation, the image of a variation $[\delta q(t)]$ of $[q(t)]$ keeping also its endpoints fixed. By definition of an NPS, the induced variations $\delta S[q_{\sast}(t)]$ and $\delta S[q(t)]$ are equal since the one of the right-hand side of~\eqref{variation} vanishes. Hence, if $[q(t)]$ leaves the action stationary, so does $[q_{\sast}(t)]$. In other words, being an NPS is, enunciated at the variational level, a sufficient condition for a transformation to map continuously the motions between themselves, that is, to be a Lie symmetry of the dynamical equations. We have here a manifestation of Curie's principle: the symmetry of the variational principle is found in its consequences, the equations of motion, and thus in the set of their solutions.

Saying that $\Phi$ is a Lie symmetry of the dynamical equations can be summarized by
\begin{align}
\mathsf E_i(L)(t_\sast,q_\sast,\dot q_\sast,\ddot q_\sast)\big|_{\{\mathsf E_j(L)(t,q,\dot q,\ddot q)=0\}}=0.\label{Lie1}
\end{align} 
Expanding this relation up to the first order in $\varepsilon$, it may be restated by the more succinct identity
\begin{align}
\boldsymbol\xi^{[2]}(\mathsf E_i(L))\big|_{\{\mathsf E_j(L)=0\}}=0\label{Lie2}
\end{align}
in $t,q,\dot q,\ddot q$. Obviously, much more can be said. Viewing $t_\sast$, $q^i_\sast(t_\sast)$, and $\dot q^i_\sast(t_\sast)$, as implicit functions of $t$ along $[q(t)]$, the two terms of~\eqref{variation} can be gathered under a single integral:
\begin{align*}
\delta_{\boldsymbol\xi}S[q(t)]=\int_{t_1}^{t_2}\bigg[L(t_\sast,q^i_\sast,\dot q^i_\sast)\frac{\rmd t_\sast}{\rmd t_{\phantom\sast}}-L(t,q^i,\dot q^i)\bigg]\rmd t.
\end{align*}
By definition, $\Phi$ is thus an NPS with BH term $\mathscr F$ iff one has, up to the first order in $\varepsilon$,
\begin{align}
L(t_\sast,q^i_\sast,\dot q^i_\sast)\frac{\rmd t_\sast}{\rmd t_{\phantom\sast}}-L(t,q^i,\dot q^i)=\varepsilon\,\frac{\rmd f}{\rmd t}\,,\label{Noether_active}
\end{align}
where $f(t,q^i)$ is the representative of $\mathscr F$ in the considered system. Then, expanding~\eqref{Noether_active} up to the first order in $\varepsilon$ yields the characterization
\begin{align}
\boldsymbol\xi^{[1]}(L)+\dot\tau L=\dot f.\label{RTidentity}
\end{align}
This identity in $t,q,\dot q$, known as \emph{Rund-Trautman identity}~\cite{Rund,Trautman1967}, is a necessary and sufficient condition for $\Phi$ to be an NPS with BH term $\mathscr F$.

Before coming to the interplay between NPS and first integrals, let us pursue further the general analysis through the passive viewpoint which is, in our opinion, the most meaningful. Consider the change of system $\{t,q^i\}\to\{t^\sast,q^{i\sast}\}$ generated by $\Phi$. Multiplying~\eqref{Noether_active} by $\rmd t/\rmd t_\sast$, one obtains the equivalent relation
\begin{align}
L(t_\sast,q^i_{\sast},\dot q_\sast^{i})-L^*(t^\sast,q^{i\sast},\mathring q^{i\sast})=\varepsilon\,\frac{\rmd f^*}{\rmd t^*}\,,\label{passive_Noether}
\end{align} 
where $f^*$ is the representative of $\mathscr F$ in the transformed system. Expression~\eqref{passive_Noether} says that the Lagrangian $L^*$ is the same function of $(t^\sast,q^{i\sast},\mathring q^{i\sast})$ than $L$ is of $(t,q^i,\dot q^i)$, up to an infinitesimal gauge term without dynamical meaning. Hence, $\Phi$ is an NPS iff the Lagrange expressions in $(t,q^i,\dot q^i,\ddot q^i)$ and $(t^\sast,q^{i\sast},\mathring q^{i\sast},\ringring q^{\,i\sast})$ are functionally the same. This demonstrates that the condition of being an NPS is much stronger than simply being a Lie symmetry. Indeed, in the passive picture, a Lie symmetry of a system of equations transforms it into another one whose solutions are functionally the same. But, for an NPS, one has utterly
\begin{align}
\mathsf E_{i*}(L^*)(t^\sast,q^{i\sast},\mathring q^{i\sast},\ringring q^{\,i\sast})=\mathsf E_i(L)(t_\sast,q_\sast^{i},\dot q_\sast^{i},\ddot q_\sast^{i}).\label{Noether1}
\end{align}
Expanding this equality up to the first order in $\varepsilon$ by using~\eqref{covariance}, it is shown to be equivalent to
\begin{align}
\boldsymbol\xi^{[2]}(\mathsf E_i(L))=-\dot\tau\,\mathsf E_i(L)-(\partial_i\xi^j-\dot q^j\partial_i\tau)\mathsf E_j(L).\label{Noether2}
\end{align}
Conditions~\eqref{Noether1} and~\eqref{Noether2} are indeed much restrictive than~\eqref{Lie1} and~\eqref{Lie2}, respectively. Moreover, after some cumbersome but straightforward algebra, it may be explicitly verified that equation~\eqref{Noether2} can be rewritten
\begin{align}
\mathsf E_i\big(\boldsymbol\xi^{[1]}(L)+\dot\tau L\big)=0\label{Noether3}
\end{align}
and is nothing else but the necessary and sufficient condition for the existence of a function $f(t,q^i)$ verifying~\eqref{RTidentity}. \textit{En r\'esum\'e}, NPS are exactly the continuous transformations leaving invariant the Lagrange expressions in the strong sense~\eqref{Noether1}.

\subsection{First integrals and invariance issues}\label{subsec:FI}

Let us introduce the \emph{Rund-Trautman expression}
\begin{align*}
\text{RT}(L,\boldsymbol\xi,f)\coloneqq\dot f-\boldsymbol\xi^{[1]}(L)-\dot\tau L
\end{align*}
whose identical vanishing is, according to~\eqref{RTidentity}, a necessary and sufficient condition for $\Phi$ to be an NPS. It can adopt the suggestive form
\begin{align}
\!\!\!\mathrm{RT}(L,\boldsymbol\xi,f)=\frac{\rmd}{\rmd t}\Big[f+H\tau-p_i\xi^i\Big]-(\xi^i-\dot q^i\tau)\mathsf E_i(L)\label{RT2}
\end{align}
where $p_i$ is the momentum conjugate to $q^i$ and $H=p_i\dot q^i-L$ the Hamiltonian (all of which are gauge dependent). The Rund-Trautman identity can thus be rewritten
\begin{align}
(\xi^i-\dot q^i\tau)\mathsf E_i(L)=\frac{\rmd}{\rmd t}\Big[f+H\tau-p_i\xi^i\Big].\label{Noether_identity}
\end{align}

In this form, it is a `divergence relation' \textit{\`a la} Noether in the specific context of classical mechanics. It is sometimes named \textit{Noether-Bassel-Hagen identity} \cite{Sundermeyer} since it is how it appears, \textit{mutatis mutandis}, in the formula (7) of Bessel-Hagen which generalizes the formula (12) of Noether. One immediately sees from~\eqref{Noether_identity} that if $\Phi$ is an NPS with BH term $f$ then the quantity
\begin{align}
I\coloneqq f+H\tau-p_i\xi^i=f-p_\mu\xi^\mu,\label{INoether}
\end{align}
in which $p_0=-H$, is a first integral\footnote{Historically, the term `first integral' referred to the equality $I=C$. Due to a semantic shift, it now designates $I$ itself.} of the problem in the sense that it keeps a constant value $C$ during the motion. The derivation of the ten well-known classical first integrals arising from the invariance under the Galilean group will not be discussed one more time here; it can be found in most textbooks [see e.g. \textcite{Logan}].\footnote{Regarding historical papers, one can also consult \textcite{Hill,Bessel-Hagen,Havas1969}.} Let us rather discuss some invariance issues of the formalism without which we would not be able to say that NPS are symmetries of variational problems.

\textit{(i) Invariance under extended coordinate transformations.} By definition, a Noether symmetry with BH term $\mathscr F$ is a property independent of the chosen extended coordinate system. This assertion can be made more precise by considering a transformation $\{t,q^i\}\to\{t',q^{i'}\}$ and verifying the obvious relation
\begin{align*}
\text{RT}(L',\boldsymbol\xi,f)=\text{RT}(L,\boldsymbol\xi,f')\,\frac{\rmd t}{\rmd t'}\,,
\end{align*}
where $L'$ is given by~\eqref{coordinate_change_L} and $f'$ is the representant of $\mathscr F$ in the new system. The first integral is obviously left invariant while keeping the same form:
\begin{align}
I=f-p_\mu\xi^\mu=f'-p_{\mu'}\xi^{\mu'}.\label{I_system}
\end{align}

\textit{(ii) Invariance under gauge transformations.} Let us denote by $\overline S$ the action built on $\overline L$ after a gauge transformation~\eqref{gauge_transformation}. Its variation is thus
\begin{align*}
\delta_{\boldsymbol\xi}\overline S[q(t)]&=\delta_{\boldsymbol\xi}S[q(t)]+\Big[\Lambda(t_\sast,q\sast^i(t_\sast))\Big]_{t_{1\sast}}^{t_{2\sast}}-\Big[\Lambda(t,q(t))\Big]_{t_1}^{t_2}\\
&=\delta_{\boldsymbol\xi}S[q(t)]+\varepsilon \Big[\boldsymbol\xi(\Lambda)\Big]_{t_1}^{t_2}.
\end{align*}
Consequently, if $\Phi$ is an NPS of the variational problem in terms of the action $S$ with BH term $f$ then it is such a symmetry in terms of $\overline S$ with BH term \cite{Leone2015}
\begin{align}
\overline f=f+\boldsymbol\xi(\Lambda).\label{transform_gauge_term}
\end{align}
Furthermore, one has
\begin{align*}
\text{RT}(\overline L,\boldsymbol\xi,\overline f)=\text{RT}(L,\boldsymbol\xi,f),
\end{align*}
and here again:
\begin{align*}
I=f-p_\mu\xi^\mu=\overline f-\overline p_\mu\xi^\mu,
\end{align*}
where the $\overline p_\mu$ are the new extended momenta
\begin{align*}
\overline p_\mu=p_\mu+\frac{\partial\Lambda}{\partial q^\mu}\,.
\end{align*}

The two invariance issues discussed above have important consequences. First of all, it is clear from~\eqref{transform_gauge_term} that if the `symmetric gauge condition'
\begin{align}
f+\boldsymbol\xi(\Lambda)=0\label{SGC}
\end{align}
is fulfilled then the symmetry is strict for $\overline L$. Now, assuming that $\overline L$ is the Lagrangian of the problem expressed in a symmetric gauge, and an adapted system $\{Q^\mu\}=\{T,Q^i\}$ which reduces $\Phi$ to a translation along $Q^\alpha$, the symmetry condition simply becomes an independence of $\overline L$ on $Q^\alpha$. Then, by~\eqref{I_system}, $I$ is (up to a sign) equal to the conjugate momentum $\overline P_\alpha$. Since the latter is built from $\overline L$, it cannot depend on $Q^\alpha$. Hence, $I$ is necessarily an invariant of the transformation:
\begin{align}
\boldsymbol\xi^{[1]}(I)=0.\label{I_invariance}
\end{align}
Note that this result was very easily derived by the use of an adapted system. Compare with the coordinate-free proof given in \textcite{Sarlet_Cantrijn}, proposition 2.2.\footnote{Obviously, one can object that coordinate-free demonstrations are conceptually more satisfying in a mathematical point of view. However, we can oppose the fact that we did not use an arbitrary system but a very peculiar one, canonically related to $\boldsymbol\xi$, which trivializes the study.} It was also present in Noether's paper, in her general context but with the same restriction to point symmetries since, as she explains, it is no more guaranteed for generalized symmetries. Some elements on that issue in classical mechanics can be found in \textcite{Sarlet_Cantrijn}.

In sum, an NPS $\Phi$ expresses the existence of an \emph{ignorable coordinate} $Q^\alpha$, in a suitable gauge and coordinate system (adapted to $\Phi$), while it maps any motion into another motion `labelled' by the same value of the first integral $I=-P_\alpha$. Using the terminology of quantum mechanics, one would say that $I$ is a `good (classical) number' of the problem.

\section{Some applications}\label{sec:applications}

\subsection{The usefulness of form invariance}

The passive viewpoint can be exploited advantageously when the Lagrangian is known to be form invariant under some class of extended coordinate transformations. Contemplate for example the standard Lagrangian
\begin{align}
L=\frac12\,g_{ij}\dot q^i\dot q^j+A_i\dot q^i-V.\label{model}
\end{align}
Whatever the problem under consideration be, it is interpretable as the Lagrangian of a unit mass particle\footnote{The `C-point' in \textcite{Lanczos}.} evolving in a space $\mathcal M$ endowed with a metric $g$ and coupled to a covector field $A$ and a scalar field $\mathscr V$. The three fields $g$, $A$, $\mathscr V$, are possibly time dependent and respectively represented by $(g_{ij})$, $(A_i)$, $V$, in the considered coordinate system. The problem is `natural' when $A$ is zero. The two fields $A$ and $\mathscr V$ are obviously not gauge invariant since any Lagrangian gauge transformation~\eqref{gauge_transformation} can be absorbed by $A$ and $V$ via a transformation
\begin{align*}
(V,A)\longrightarrow(V-\partial_t\Lambda,A_i+\partial_i\Lambda).
\end{align*}
The Lagrange expressions of~\eqref{model} are
\begin{align*}
\mathsf E_i(L)=-g_{ij}\ddot q^j-\Gamma_{ijk}\,\dot q^j\dot q^k+\big(B_{ij}-\partial_t g_{ij}\big)\dot q^j+E_i
\end{align*}
where the quantities
\begin{align*}
\Gamma_{ijk}\coloneqq\frac12\big(\partial_ig_{kj}+\partial_jg_{ik}-\partial_kg_{ij}\big)
\end{align*}
are the Christoffel symbols of the first kind, whereas
\begin{align*}
B_{ij}\coloneqq\partial_iA_j-\partial_jA_i\quad\text{and}\quad E_i\coloneqq-\partial_iV-\partial_tA_i
\end{align*}
are the components of gauge invariant fields $B$ and $E$ built from $A$ and $\mathscr V$.

It is clear that $L$ is form invariant under the whole class of time-independent coordinate transformations unaffecting the time, i.e.
\begin{align}
(t,q^i)\longrightarrow(t,q^{i'}=\phi^i(q^k)).\label{type1}
\end{align}
In fact, the Lagrangian has the property of `manifest covariance' with respect to such transformations inasmuch as its expression in the new system is
\begin{align*}
L'=\frac12\,g_{i'j'}\,\dot q^{i'}\dot q^{j'}+A_{i'}\dot q^{i'}-V'
\end{align*}
where $(g_{i'j'})$, $(A_{i'})$, and $V'$ are the representatives of $g$, $A$, and $\mathscr V$, in the new system.  

In particular, a continuous transformation $\Phi$ of the type~\eqref{type1} is generated by a vector field $\boldsymbol\xi=(0,\xi^i)$ with time-independent components $\xi^i$, which can be seen as a vector field in $\mathcal M$. The NPS characterization~\eqref{passive_Noether} is here an equality between two polynomial expressions in the velocities. It is verified iff each of their monomials of the same kind coincide, leading to the three sets of identities
\begin{equation}
\begin{aligned}
g_{ij}(t,q^{k}_\sast)-g_{i\sast j\sast}(t,q^{k\sast})&=0,\\
A_{i}(t,q^{k}_\sast)-A_{i\sast}(t,q^{k\sast})&=\varepsilon\,\partial_{i\sast}f^*(t,q^{k\sast}),\\
V(t,q^{k}_\sast)-V^*(t,q^{k\sast})&=-\varepsilon\,\partial_tf^*(t,q^{k\sast}),\label{modelinvariance}
\end{aligned}
\end{equation}
where $(g_{i\sast j\sast})$, $(A_{i\sast})$, and $V^*$ are once again the representatives of $g$, $A$, and $\mathscr V$, in the transformed system $\{t^\sast,q^{i\sast}\}=\{t,q^{i\sast}\}$. Obviously, the expressions $\mathsf E_i(L)$ have the same property of form invariance and the characterization~\eqref{Noether1} leads to the four sets of identities 
\begin{align*}
g_{i\sast j\sast}(t,q^{l\sast})&=g_{ij}(t,q^{l}_\sast),\\
\Gamma_{i\sast j\sast k\sast}(t,q^{l\sast})&=\Gamma_{ijk}(t,q^{l}_\sast),\\
B_{i\sast j\sast}(t,q^{l\sast})-\partial_t g_{i\sast j\sast}(t,q^{l\sast})&=B_{ij}(t,q^{l}_\sast)-\partial_t g_{ij}(t,q^{l}_\sast),\\
E_{i\sast}(t,q^{l\sast})&=E_i(t,q^{l}_\sast),
\end{align*}
with the same convention whereby starred indices refers to representatives in $\{t,q^{i\sast}\}$. Taking the derivatives of the first line, it is clear that this system is equivalent to
\begin{equation}
\begin{aligned}
g_{ij}(t,q^{k}_\sast)-g_{i\sast j\sast}(t,q^{k\sast})&=0,\\
B_{ij}(t,q^{k}_\sast)-B_{i\sast j\sast}(t,q^{k\sast})&=0,\\
E_i(t,q^{k}_\sast)-E_{i\sast}(t,q^{k\sast})&=0.\label{modelEi}
\end{aligned}
\end{equation}

One recognizes in the left-hand sides of~\eqref{modelinvariance} and~\eqref{modelEi} the Lie differentials of the various fields along $(\xi^i)$ in $\mathcal M$, in its passive interpretation which is perhaps the most familiar to physicists. The Lie differential\footnote{Let $T$ be a tensor field over $\mathcal M$ whose components are $T^{i\dots}_{j\dots}$ in an arbitrary coordinate system, at a given instant, say. Its Lie derivative along $\boldsymbol\xi$ is the quantity $\mathcal L_{\boldsymbol\xi}T$ whose components $\mathcal L_{\boldsymbol\xi}T^{i\dots}_{j\dots}$ in the same coordinate system are given by the rule
\begin{align}
T^{i\dots}_{j\dots}(q^k_\sast)-T^{i\sast\dots}_{j\sast\dots}(q^{k\sast})=\varepsilon\,\mathcal L_{\boldsymbol\xi}T^{i\dots}_{j\dots}(q^{k}_\sast)+\mathrm{o}(\varepsilon),\label{Lie_derivative}
\end{align}
through the passive picture. It can be shown that $\mathcal L_{\boldsymbol\xi}T$ is a tensor of the same kind than $T$. The definition given here seems to depart from the somewhat usual rule whereby a variation is modelled on the scheme `transformed quantity minus original one' but is actually more consistent. It coincides with the `dragging along' (\textit{Mitschleppen}) construction of Schouten and van Kampen \cite{Schouten1934,Schouten1954} which was later formulated in intrinsic terms by mathematicians [see e.g. \textcite{Choquet-DeWitt}]. The left-hand side of~\eqref{Lie_derivative} is the Lie differential (\textit{Liesche Differential}) in the terminology of Schouten and van Kampen. Surprisingly enough, the Lie derivative in the quite recent book \textcite{Petrov} is defined as the opposite of the Lie differential. Using~\eqref{Lie_derivative} in conjunction with the transformation laws of the representatives of tensor fields, one obtains easily
\begin{align*}
\mathcal L_{\boldsymbol\xi}\phi=\boldsymbol\xi(\phi)
\end{align*}
for a scalar field,
\begin{align*}
\mathcal L_{\boldsymbol\xi}T_i=\boldsymbol\xi(T_i)+T_k\partial_i\xi^k
\end{align*}
for a covector field,
\begin{align*}
\mathcal L_{\boldsymbol\xi}T_{ij}=\boldsymbol\xi(T_{ij})+T_{kj}\partial_i\xi^k+T_{ik}\partial_j\xi^k
\end{align*}
for a covariant tensor field of rank 2, etc. One can derive a general expression for a tensor of arbitrary kind and for even more general quantities \cite{Yano} but the three formulas above suffice for our context. The Lie derivative of any of those quantities always contains a term evaluating the rate of change of its components in $\boldsymbol\xi$'s direction. If the convention `transformed minus original' were retained in~\eqref{Lie_derivative}, this term would become an evaluation in the direction of $-\boldsymbol\xi$ which is less satisfying.} is the tool which measures how fields transform \cite{Yano} and the last system means, in the passive picture, that $\Phi$ is an NPS iff the three background fields $g$, $B$, and $E$, are seen by the observer $\mathcal O^*$ exactly as they appear to $\mathcal O$. It corresponds, in the active viewpoint, to the invariance of the distributions of these fields along the direction of the transformation. 

One knows from our general considerations on NPS that the two systems~\eqref{modelinvariance} and~\eqref{modelEi} are equivalent. Here, the equivalence is obvious because the two last lines of~\eqref{modelinvariance} are nothing else but the requirement that the couple of fields $(V,A)$ is seen by $\mathcal O^*$ just like it appears to $\mathcal O$, up to a meaningless gauge transformation \cite{Forgacs1980}.

The invariance of $g$ in~\eqref{modelinvariance} or~\eqref{modelEi} notably expresses the fact that $\Phi$ is a continuous isometry, or to put it another way, that $\boldsymbol\xi$ is a Killing vector field of the metric. Applying the formula of the Lie derivative, this invariance amounts to the verification of the so-called \emph{Killing equations} 
\begin{align*}
\mathcal L_{\boldsymbol\xi}g_{ij}=\boldsymbol\xi(g_{ij})+g_{kj}\partial_i\xi^k+g_{ik}\partial_j\xi^k=0.
\end{align*}
In the particular case of the free particle, it is the only condition to fulfil.\footnote{It can easily be shown that there are at most $n(n+1)/2$ independent Killing vector fields in an $n$-dimensional space \cite{Wald}. They form the Lie algebra of the isometry group of the space. The usual three-dimensional Euclidean space and Minkowski space, for example, are maximally symmetric since they admit isometry groups of dimensions $6=3\cdot 4/2$ (translations and rotations) and $10=4\cdot 5/2$ (Poincar\'e transformations), respectively.} This is the reason why the Rund-Trautman identity is also often called \emph{generalized Killing equation} \cite{Logan,Vujanovic} or \emph{Killing-type equation} \cite{Sarlet_Cantrijn}. 

The argument of form invariance can be used for the wider class of extended coordinate transformations
\begin{align}
(t,q^i)\longrightarrow(t'=\phi^0(t),q^{i'}=\phi^i(t,q^k)).\label{type2}
\end{align}
Indeed, albeit non manifestly covariant, the new Lagrangian adopts the same form:
\begin{align}
L'=\frac12\,\tilde g_{i'j'}\mathring q^{i'}\mathring q^{j'}+\widetilde A_{i'}\mathring q^{i'}-\widetilde V',\label{same_form}
\end{align}
with
\begin{align*}
\tilde g_{i'j'}&=g_{kl}\,\frac{\partial q^k}{\partial q^{i'}}\frac{\partial q^l}{\partial q^{j'}}\frac{\rmd t'}{\rmd t}\,,\\
\widetilde A_{i'}&=A_j\,\frac{\partial q^j}{\partial q^{i'}}+g_{jk}\,\frac{\partial q^j}{\partial q^{i'}}\frac{\partial q^k}{\partial t'}\frac{\rmd t'}{\rmd t}\\
\widetilde V'&=V\frac{\rmd t}{\rmd t'}-A_i\frac{\partial q^i}{\partial t'}-\frac12\,g_{ij}\,\frac{\partial q^i}{\partial t'}\frac{\partial q^j}{\partial t'}\frac{\rmd t'}{\rmd t}\,.
\end{align*}
A continuous transformation of the type~\eqref{type2} has a generator $\boldsymbol\xi=(\tau,\xi^i)$ characterized by a component $\tau$ depending on $t$ only. The NPS characterization~\eqref{passive_Noether} is now
\begin{equation}
\begin{aligned}
g_{ij}(t_\sast,q^{k}_\sast)-\tilde g_{i\sast j\sast}(t^\sast,q^{k\sast})&=0,\\
A_{i}(t_\sast,q^{k}_\sast)-\widetilde A_{i\sast}(t^\sast,q^{k\sast})&=\varepsilon\,\partial_{i\sast}f^*(t^\sast,q^{k\sast}),\\
V(t_\sast,q^{k}_\sast)-\widetilde V^*(t^\sast,q^{k\sast})&=-\varepsilon\,\partial_{0\sast}f^*(t^\sast,q^{k\sast}).\label{modelinvariance2}
\end{aligned}
\end{equation}
These equalities obviously contain~\eqref{modelinvariance} as a special case. After an expansion up to the first order in $\varepsilon$, the above system amounts to
\begin{equation}
\begin{aligned}
\boldsymbol\xi(g_{ij})+g_{ik}\partial_j\xi^k+g_{kj}\partial_i\xi^k&=\dot\tau g_{ij},\\
\boldsymbol\xi(A_i)+A_k\partial_i\xi^k+g_{ik}\partial_t\xi^k&=\partial_if,\\
\boldsymbol\xi(V)+\dot\tau V-A_k\partial_t\xi^k&=-\partial_tf.
\end{aligned}\label{form_invariancefinal}
\end{equation}
A similar analysis based on the characterization~\eqref{Noether1} would have led to the equivalent system
\begin{align*}
\boldsymbol\xi(g_{ij})+g_{ik}\partial_j\xi^k+g_{kj}\partial_i\xi^k&=\dot\tau g_{ij},\\
\boldsymbol\xi(B_{ij})+B_{ik}\partial_j\xi^k+B_{kj}\partial_i\xi^k&=\partial_j(g_{ik}\partial_t\xi^k)-\partial_i(g_{jk}\partial_t\xi^k),\\
\boldsymbol\xi(E_i)+E_k\partial_i\xi^k+\dot\tau E_i&=B_{ki}\partial_t\xi^k,
\end{align*}
which can most simply be obtained by taking the cross derivatives of the two last lines of~\eqref{form_invariancefinal}.

In fact, the transformations~\eqref{type2} constitute the most general class of transformations leading to a new Lagrangian of the same form~\eqref{same_form}. Indeed, if $\phi^0$ depends also on the coordinates $q^i$ then the kinetic part of $L$ multiplied by $\rmd t/\rmd t'$ is a rational function of the new velocities. An infinitesimal transformation of this type makes appear in the Lagrangian a cubic term in the velocities and can never verify the identity~\eqref{passive_Noether}. Equations~\eqref{form_invariancefinal}, together with the restriction $\tau=\tau(t)$, are consequently the necessary and sufficient conditions for a continuous transformation to be an NPS of the considered problem. The direct application of the Rund-Trautman identity would have obviously lead to the same conclusion.

The method used to arrive at the determining equations through the lens of form invariance in the passive viewpoint may be deemed too tedious. Indeed, the steps from~\eqref{modelinvariance2} to~\eqref{form_invariancefinal} need a careful application of the passive transformation laws of field components or the knowledge of the Lie derivative. However, it has the merit of narrowing the investigation by exploiting the form invariance --- which is a symmetry \textit{per se} --- as a necessary condition for the functional invariance. In our opinion, it is a more profound approach than the systematic method based on the Rund-Trautman identity, especially in the case of transformations of the type~\eqref{type1}.

\subsection{Application to one-dimensional problems}

As a case study, let us focus on the rectilinear dynamics of a unit-mass particle experiencing a potential $V$, and governed by the standard Lagrangian
\begin{align*}
L=\frac12\,\dot q^2-V(t,q).
\end{align*}
Our aim is to find the potentials for which the problem admits an NPS $\boldsymbol\xi=(\tau(t),\xi(t,q))$. Applying~\eqref{form_invariancefinal}, one deduces the following characterization:
\begin{align*}
2\partial_q\xi-\dot\tau&=0,\\
\partial_t\xi-\partial_qf&=0,\\
\xi\,\partial_qV+\partial_t(\tau V)+\partial_tf&=0.
\end{align*}
No matter what the potential is, the two first lines impose to $\xi$ and $f$ the following forms
\begin{align*}
\xi(t,q)&=\frac12\,\dot\tau(t)\,q+\psi(t)\;;\\
f(t,q)&=\frac14\,\ddot\tau(t)\,q^2+\dot\psi(t)\,q+\chi(t),
\end{align*}
where $\psi$ and $\chi$ are thus far undetermined. Finally, the remaining equality constitutes a compatibility equation between $\tau$, $\psi$, $\chi$, and the potential. Bearing in mind that the functions $\tau$ and $\psi$ cannot be both identically zero otherwise the transformation is the identity and the symmetry trivial, there are two distinct cases to consider, depending on whether the time is left invariant ($\tau=0$) or not ($\tau\ne0$).

\subsubsection{The case $\tau=0$}
Here, the compatibility equation reduces to
\begin{align}
\psi(t)\,\frac{\partial V}{\partial q}+\ddot\psi(t)\,q+\dot\chi(t)=0.\label{RTVpsi}
\end{align}
Hence, apart from an irrelevant term of $t$ alone, the most general potential admitting such a symmetry has the form
\begin{align}
V(t,q)=\frac12\,a(t)q^2+b(t)q.\label{Vpsi}
\end{align}
The function $\psi$ is submitted to the differential equation
\begin{align}
\ddot\psi(t)+a(t)\psi(t)=0,\label{eqpsi}
\end{align}
whereas $\chi$ is adjusted to cancel the terms of the time alone in~\eqref{RTVpsi}:
\begin{align*}
\chi(t)=-\int b(t)\psi(t)\,\rmd t.
\end{align*}

In particular, the spatial translation is an NPS when $V$ depends linearly on $q$, i.e. when the particle is submitted to a uniform force field. Once fixed a nonzero function $\psi$ verifying~\eqref{eqpsi}, the gauge condition is fulfilled by
\begin{align*}
\Lambda(t,q)=-\frac{1}{\psi(t)}\bigg(\frac12\,\dot\psi(t)q^2+\chi(t)q\bigg).
\end{align*}
Since $t$ is left invariant, the most natural adapted system is formed by the time $t$ and the coordinate $Q=q/\psi(t)$ along which the translation is done. It leads to the symmetric Lagrangian
\begin{align*}
\overline L=\frac12\,\psi(t)^2\dot Q^2-\chi(t)\dot Q
\end{align*}
independent of $Q$. Actually, we are dealing here with a symmetry stemming only from the linearity of the dynamical equation
\begin{align*}
\ddot q+a(t)q+b(t)=0.
\end{align*}
It has been dubbed `linearity symmetry' in a recent article \cite{lienearity}. The above reduction in terms of the new coordinate $Q$ is nothing else but the Lagrangian counterpart of the usual reduction technique of linear differential equations once a nonzero solution $\psi(t)$ of their associated homogeneous equation is known. The first integral is precisely the reduced equation:
\begin{align*}
I=\dot\psi(t)q-\dot q\psi(t)+\chi(t)=-\psi(t)^2\dot Q+\chi(t)=C
\end{align*} 
and one recognizes the Wronskian between $q$ and $\psi$ when $b(t)=0$. Since the solution space of~\eqref{eqpsi} is of dimension 2, there are two independent linearity symmetries.

\subsubsection{The case $\tau\ne0$}
Changing the transformation to its inverse if necessary, one can suppose $\tau$ positive. Multiplying the compatibility equation by $\tau$, one has
\begin{align}
\boldsymbol\xi(V\tau)+\tau\,\partial_tf=0.\label{RTtau}
\end{align}
It is easily seen that if one introduces the system $(T,Q)$ given by
\begin{align*}
T=\int\frac{\rmd t}{\tau}\quad\text{and}\quad Q=\frac{q}{\sqrt\tau}-\int\frac{\psi}{\tau^{3/2}}\,\rmd t
\end{align*}
then $\boldsymbol\xi$ reduces to $\partial_T$. For later convenience, let us introduce the functions
\begin{align*}
\rho(t)=\sqrt\tau\quad,\quad \alpha(t)=\int\frac{\psi}{\tau^{3/2}}\,\rmd t\quad\text{and}\quad\beta(t)=\frac{\chi}{\tau}\,.
\end{align*}
After some lengthy but straightforward computations, one finds that
\begin{align*}
\tau\,\frac{\partial f}{\partial t}=\frac{\partial}{\partial T}\bigg(\frac12\,\ddot\rho\rho q^2+\frac{\rmd}{\rmd t}\big(\rho^2\dot \alpha\big)\rho q-\frac12\,\rho^4\dot a^2+\rho^2\beta\bigg)
\end{align*}
and~\eqref{RTtau} now reads
\begin{align*}
\frac{\partial}{\partial T}\bigg(\rho^2V+\frac12\,\ddot\rho\rho q^2+\frac{\rmd}{\rmd t}\big(\rho^2\dot \alpha\big)\rho q-\frac12\,\rho^4\dot \alpha^2+\rho^2\beta\bigg)=0.
\end{align*}
One concludes that the most general potential for which the Lagrangian admits an NPS transforming the time has the form
\begin{align}
V=\frac{1}{\rho^2}\,W\bigg(\frac{q}{\rho}-\alpha\bigg)-\frac{\ddot\rho}{2\rho}\, q^2-\frac{1}{\rho}\frac{\rmd}{\rmd t}\big(\rho^2\dot \alpha\big)q+\frac{1}{2}\,\rho^2\dot\alpha^2-\beta.\label{forme_V}
\end{align}
The gauge condition is then fulfilled by
\begin{align*}
\Lambda(t,q)=-\frac{\dot\rho}{2\rho}\,q^2-\rho\dot \alpha q+\int\big(\rho^2\dot\alpha^2-\beta\big)\rmd t.
\end{align*}
It leads, in the adapted system, to the symmetric Lagrangian
\begin{align*}
\overline L=\frac12\,\mathring Q^2-W(Q),
\end{align*}
where one has denoted the total $T$-derivative by an empty bullet. Hence, in the adapted system, the dynamics becomes derived from the conservative potential $W$, and the Hamiltonian
\begin{align*}
\overline H=\frac12\,\mathring Q^2+W(Q)=\frac12\,\big(\rho\dot q-\dot\rho q-\rho^2\dot\alpha\big)^2+W\bigg(\frac{q}{\rho}-\alpha\bigg)
\end{align*}
is the first integral $I$ generated by the symmetry. Our results are in accordance with the conclusions of Lewis and Leach \cite{Lewis-Leach} who used another approach based on the Poisson bracket.

Each distinct decomposition of a given potential $V$ into the form~\eqref{forme_V}, if any, amounts to a symmetry. Clearly, if $V$ does not fit~\eqref{Vpsi}, there is at most one independent NPS. Otherwise, up to an unimportant constant which can be incorporated into $\beta$, the most general function $W$ in a position to enter the decomposition of a potential~\eqref{Vpsi} is
\begin{align*}
W(Q)=\frac12\,AQ^2+BQ,
\end{align*}
where $A$ and $B$ are arbitrary coefficients. Then, $\rho$ and $\alpha$ must verify
\begin{align}
A&=\rho^3(\ddot\rho+a\rho),\label{eqA}\\
B&=\rho^2\frac{\rmd}{\rmd t}\big(\rho^2\dot\alpha\big)+\rho^3b+A\alpha,\label{eqB}
\end{align}
while $\beta$ is adjusted to cancel the terms of $t$ alone in~\eqref{forme_V}. Since $A$ and $B$ are arbitrary, equations~\eqref{eqA} and~\eqref{eqB} amount to the identical vanishing of the derivative of their right-hand side, that is
\begin{align*}
\frac14\,\dddot\tau+a\dot\tau+\frac12\,\dot a\tau&=0,\tag{42'}\\
\ddot\psi+a\psi+\frac32\,b\dot\tau+\dot b\tau&=0\tag{43'}.
\end{align*}
Actually, these two equations are the necessary and sufficient conditions on $\tau$ and $\psi$ that we obtain when the potential~\eqref{Vpsi} is injected in the compatibility equation. Besides the linearity symmetries which were already found when $\tau$ was set to zero, one obtains three supplementary independent NPS: as many as the order of~(\ref{eqA}\textcolor{red}{'}). 

The coefficients $A$ and $B$ are integrating constants of~(\ref{eqA}\textcolor{red}{'}) and ~(\ref{eqB}\textcolor{red}{'}). By tuning their values, the initial problem can be mapped into the one of a free particle ($A=B=0$), a particle immerged in a static uniform force field ($A=0$), or a time-independent harmonic oscillator ($A>0$). In particular, the most interesting case of the time-dependent harmonic oscillator, for which $a(t)$ is the square of the frequency $\omega(t)$ while $b(t)$ is zero, can be mapped to the one of an oscillator with unit frequency. It suffices to set $\alpha=\psi=\chi=0$ and to find a solution $\rho$ to the so-called Ermakov equation
\begin{align*}
\ddot\rho+\omega^2(t)\rho=\rho^{-3}.
\end{align*}
The first integral $I=\overline H$ thereby obtained is the Ermakov-Lewis invariant \cite{Ermakov,Lewis}.

\subsection{A manifestation of Noether's second theorem: the parametrization invariance}\label{subsec:parametrization}

\subsubsection{The parametrization invariance}

Let us determine the conditions for which a variational problem based on a Lagrangian $L(t,q^i,\dot q^i)$ is parametrization-invariant in the following sense: $L$ is left invariant by any transformation of the form
\begin{align}
(t,q^i)\longrightarrow(t'=\phi(t,q^k),q^i).\label{transformation_time}
\end{align} 
Alternatively stated, a parametrization-invariant formulation is a formulation for which the parameter can be chosen arbitrarily without incidence on the functional form of the equations. Such a request will have important consequences. Indeed, suppose that $[q^i(t)]$ is a solution of a problem having the sought property. By symmetry, the actively transformed evolution $[q'^i(t)]$ given by $q'^i(t')=q^i(\phi(t,q^k(t)))$ will also be a solution whatever our choice of $\phi$ be. But there is in particular an infinity of different ways of choosing $\phi$ so that $[q'^i(t)]$ obeys to the same initial conditions than $[q^i(t)]$. Hence, for any initial conditions, the request leads inevitably to an infinite set of solutions. In a Newtonian point of view for which $t$ is an absolute time, this situation would severely violate the determinism unless the variational problem is assorted with auxiliary conditions allowing to `recover' the time. To put it differently, in a parametrization-invariant formulation, the arbitrariness in the choice of the parameter implies its insignificance: it must be understood as an ingredient without meaning \textit{a priori}\footnote{One can easily understand the trouble in the mind of physicists about the general invariance which implies the insignificance of the coordinates, these common objects which were before always charged of (metrical) meaning. It is well illustrated by Einstein himself in the following excerpt of his autobiographical notes where he explains why it took seven years between the idea of generalizing his theory of relativity (1908) and its realization \cite{Schilpp}:
\begin{quote}
\textit{Warum brauchte es weiterer 7 Jahre f\"ur die Aufstellung der allgemeinen Rel. Theorie? Der haupt\-s\"achliche Grund liegt darin, dass man sich nicht so leicht von der Auffassung befreit, dass den Koordinaten eine unmittelbare metrische Bedeutung zukommen m\"usse.}
\end{quote}
According to Schilpp's translation:
\begin{quote}
Why were another seven years required for the construction of the general theory of relativity? The main reason lies in the fact that it is not so easy to free oneself from the idea that co-ordinates must have an immediate
metrical meaning.
\end{quote}
The terminology `world parameters' (\textit{Weltparameters}) used by Hilbert to name arbitrary coordinates is on this aspect well adapted \cite{Janssen}.}; it is only through our choice \textit{a posteriori} that it acquires a `reality'.

The lack of determinism evoked above can be rephrased as the impossibility of putting the Euler-Lagrange equations in the normal form $\ddot q^i=\Omega^i(t,q^k,\dot q^k)$. Since one has
\begin{align*}
\mathsf E_i(L)=\frac{\partial L}{\partial q^i}-\frac{\partial^2L}{\partial\dot q^i\partial t}-\dot q^j\,\frac{\partial^2L}{\partial\dot q^i\partial  q^j}-\ddot q^{\,j}\,\frac{\partial^2L}{\partial\dot q^i\partial\dot q^j}\,,
\end{align*}
it means that the Lagrangian is certainly singular in the sense that its Hessian matrix with respect to the velocities,
\begin{align*}
\mathsf H\coloneqq\bigg(\frac{\partial^2L}{\partial\dot q^i\partial\dot q^j}\bigg)_{ij},
\end{align*}
is singular. Among the consequences, we are unable to express unambiguously the velocities as functions of the time, the coordinates, and the momenta. Hence, no Legendre transform to pass from the Lagrangian to an Hamiltonian is allowed in the usual sense, that is, without having recourse to the theory of Dirac constraints \cite{Dirac}. 
 
Let us now pursue the symmetry analysis by considering the continuous transformations of the type~\eqref{transformation_time}. They are generated by all the fields of the form $\tau(t,q^i)\partial_t$ and must be strict NPS. For any evolution $[q(t)]$ between two instants $t_1$ and $t_2$, Noether's identity~\eqref{Noether_identity} implies
\begin{align*}
-\int_{t_1}^{t_2}\tau\,\dot q^i\mathsf E_i(L)\,\rmd t=\Big[H\tau\Big]_{t_1}^{t_2}.
\end{align*}
This equality must in particular be true for all functions $\tau(t)$ vanishing at the extremities of time. Hence, by the fundamental lemma of the calculus of variations, one deduces the identity
\begin{align}
\dot q^i\,\mathsf E_i(L)=0.\label{constraint}
\end{align}
It is actually a manifestation of the second Noether's theorem: the infinite symmetry group generated by all the vector fields $\tau(t,q^i)\partial_t$ has for consequence a dependency relationship between the Euler-Lagrange expressions. It follows from~\eqref{constraint} that the system of Euler-Lagrange equations is underdetermined: one of them being redundant, the $n$ degrees of freedom outnumber the independent equations. Furthermore, since $\tau$ is arbitrary, the Rund-Trautman identity
\begin{align*}
\boldsymbol{\xi}^{[1]}(L)+\dot\tau L=\tau\,\partial_tL-\dot\tau H=0
\end{align*}
imposes the two subsequent identities, derived \textit{\`a la} Klein,
\begin{align}
\partial_tL=0\quad\text{and}\quad H=0,\label{KU}
\end{align}
known as the \emph{Zermelo conditions} \cite{Bolza1904}. The second one amounts to the homogeneity of degree 1 of $L$ in the velocities. Reciprocally, it is quite clear that these conditions are also sufficient because then, for any transformation~\eqref{transformation_time}, one has
\begin{align*}
L(q^i,\dot q'^i)=L\bigg(q^i,\dot q^i\,\frac{\rmd t}{\rmd t'}\bigg)=L(q^i,\dot q^i)\frac{\rmd t}{\rmd t'}=L'(q^i,\mathring q^i),
\end{align*}
where the empty bullet symbolizes the total $t'$-derivative.

\subsubsection{An Application to extended Lagrangians}

Reconsider a general variational problem as discussed in section~\ref{sec:Noether} and let us introduce an extra variable $\sigma$ which is supposed to strictly increase with $t$. Then, contemplate the function
\begin{align*}
\mathscr L(q^\mu,v^\mu)=L(t,q^i,\dot q^i)\frac{\rmd t}{\rmd\sigma}=L\bigg(q^0,q^i,\frac{v^i}{v^0}\bigg)v^0,
\end{align*}
where $v^\mu$ designates the total derivative of $q^\mu$ with respect to $\sigma$. It is easily verified that
\begin{align}
\frac{\partial\mathscr L}{\partial q^\mu}=v^0\,\frac{\partial L}{\partial q^\mu}\quad\text{and}\quad \frac{\partial\mathscr L}{\partial v^\mu}=p_\mu\,.\label{derivatives_mu}
\end{align}
Seeing $q^0$ as a coordinate on equal footing with the others, the $n+1$ Lagrange expressions of $\mathscr L$ are the quantities $\mathscr E_\mu(\mathscr L)$ where
\begin{align*}
\mathscr E_\mu=\frac{\partial}{\partial q^\mu}-\frac{\rmd}{\rmd\sigma}\frac{\partial}{\partial v^\mu}\,.
\end{align*}
A direct application of~\eqref{derivatives_mu} shows that the expressions $\mathscr E_i(L)$ and $\mathsf E_i(L)$ are mutually related by
\begin{align}
\mathscr E_i(\mathscr L)=v^0\mathsf E_i(L).\label{EL_extended}
\end{align}
By construction, $\mathscr L$ is parametrization-independent, thus singular, since $\sigma$ is insignificant. Hence, the following identity
\begin{align*}
v^\mu\mathscr E_\mu(\mathscr L)=0
\end{align*}
holds and one deduces the relationship
\begin{align}
\mathscr E_0(\mathscr L)=-v^i\mathsf E_i(L)\label{EL_time}
\end{align}
which might also have been derived through another use of~\eqref{derivatives_mu}, but in a less straightforward way. 

The relationships~\eqref{EL_extended} demonstrate that the Euler-Lagrange equations of $\mathscr L$ with respect to the coordinates $q^i$ amount to the equations of motion. Since their solutions automatically cancel out the redundant expression~\eqref{EL_time}, the problem may equivalently be addressed in terms of the \emph{extended Lagrangian} $\mathscr L$: we are faced with Weierstrass' parametric representation of the same problem \cite{Bolza1904}. The latter is not only a mere `curiosity'. Beyond the physical relevance of $\mathscr L$ in the `passage' \cite{Dirac} from Newtonian mechanics, where an absolute time exists, to (non Galilean) relativistic theories, where no such time exists, this object can also be of great utility, even in classical mechanics. The next section will provide a fruitful application of $\mathscr L$. For the time being, let us demonstrate its usefulness to establish formula~\eqref{covariance}. It is a basic fact of Lagrangian mechanics that under a coordinate transformation $\{q^i\}\to\{q^{i'}=\phi^i(t,q^k)\}$ unaffecting the time, the Lagrange expressions transform covariantly:
\begin{align}
\mathsf E_{i'}(L')=\frac{\partial q^j}{\partial q^{i'}}\,\mathsf E_j(L).\label{covariancesimple}
\end{align}
Now, consider an arbitrary extended coordinate transformations $\{q^\mu\}\to\{q^{\mu'}\}$ as well as the extended Lagrangians $\mathscr L$ and $\mathscr L'$ constructed from $L$ and $L'$. All of them are related by
\begin{align*}
\mathscr L'=L'v^{0'}=Lv^0=\mathscr L,
\end{align*}
where $v^{\mu'}$ designates the derivative of $q^{\mu'}$ with respect to $\sigma$, and one has the relation of covariance
\begin{align*}
\mathscr E_{\mu'}(\mathscr L')=\frac{\partial q^\nu}{\partial q^{\mu'}}\,\mathscr E_\nu(\mathscr L).
\end{align*} 
In particular:
\begin{align*}
\mathscr E_{i'}(\mathscr L')=\frac{\partial q^0}{\partial q^{i'}}\,\mathscr E_0(\mathscr L)+\frac{\partial q^j}{\partial q^{i'}}\,\mathscr E_j(\mathscr L).
\end{align*}
Then, from~\eqref{EL_extended} and~\eqref{EL_time}, one deduces
\begin{align*}
v^{0'}\mathsf E_{i'}(L')=\bigg(\frac{\partial q^j}{\partial q^{i'}}-\frac{\partial q^0}{\partial q^{i'}}\frac{v^j}{v^0}\bigg)v^0\mathsf E_j(L).
\end{align*}
Dividing by $v^{0'}$ produces formula~\eqref{covariance}.

\section{Noether point symmetries and Routh reduction}\label{sec:Routh}

\subsection{The Routh reduction and its usefulness}

In an essay on the stability of motion, Routh \cite{Routh1877} introduced a recipe to eliminate from the very beginning the ignorable coordinates\footnote{Since they do not appear in the Lagrangian, Routh called them `absent coordinates'.} in a problem via a `modification' of its initial Lagrangian. This method is often referred to as the \emph{ignoration of coordinates} after the terminology that Thomson and Tait introduced in the revised edition of their treatise on natural philosophy~\cite{TT}. However, what these authors called in this way was actually a similar elimination but realized at the level of the kinetic energy specifically, for those systems characterized by the existence of some coordinates, said `cyclic', which do not appear in the kinetic energy whereas no (generalized) force acts in their direction.\footnote{This is how Lamb \cite{Lamb} defines `cyclic systems'. Regarding coordinates, the adjective `cyclic' is nowadays a synonym for `ignorable' and its use was considered as a `pity' by Synge because of its confusion with the topological sense of the term \cite{Synge1960}.} 
As was noticed by Lamb and Pars~\cite{Pars}, Larmor~\cite{Larmor} gave the first a variational version of Routh's procedure. For the sake of completeness, we give a brief account of the method.

Suppose that the Lagrangian expressed in a system $\{t,q^i\}$ admits exactly $m<n$ ignorable coordinates, $q^1,\dots,q^m$ say. The motion is thus submitted to the $m$ constraints $p_i=C_i$ where $C_i$ is the actual constant value of $p_i$ ($i=1,\dots, m$). However, this information is not taken into account in the original Hamilton's principle which considers all the evolutions between two endpoints, and \textit{a fortiori} the irrelevant ones which do not respect the constraints. Narrowing the study to evolutions compatible with the constraints leads to the well-known reduced principle \cite{Larmor}
\begin{align}
\delta\int \big(L-p_1\dot q^1-\dots- p_m\dot q^m\big)\,\rmd t=0,\label{redvar}
\end{align}
for arbitrary variations of $q^{m+1},\dots,q^n$ vanishing at the extremities of time while the variations of $q^1,\dots,q^m$ are adapted for the sole purpose of maintaining the constraints. In fact, it is assumed that the equalities
\begin{align*}
\frac{\partial L}{\partial\dot q^i}=C_i\qquad(i=1,\dots, m)
\end{align*}
determine unambiguously $\dot q^1,\dots,\dot q^m$ as functions
\begin{align}
\dot q^i=\varphi^i(t,q^{m+1},\dots,q^n,\dot q^{m+1},\dots,\dot q^n,C_1,\dots,C_m)\label{phi}
\end{align}
but it is certainly the case if $L$ is regular, an assumption which will be tacitly understood. Hence, the function
\begin{align}
R\coloneqq L-C_1\dot q^1-\dots-C_m\dot q^m \label{Routhian}
\end{align}
in which each occurrence of $\dot q^i$ is replaced by $\varphi^i$, for $i=1,\dots, m$, is a genuine Lagrangian governing the dynamics of the $n-m$ last degrees of freedom. Once the latter solved, the ignored ones are obtained by a quadrature based on~\eqref{phi}.\footnote{Considering a dynamical system of equations $\dot x^i=X^i(t,x^k)$, there are two main ways of reducing its order by one. The first one supposes the existence of a first integral $I(t,x^k)=C$ and consists in performing a change of variables $x^i\to y^i$ such that $y^1=I$, say. The second one supposes that $X^2,\dots,X^N$ do not depend on $x^1$, in which case the system reduces to $\dot x^i=X^i(t,x^2,\dots,x^N)$, $i=2,\dots,N$, while $x^1$ is obtained by a final quadrature. The existence of an NPS, and thus of an ignorable coordinate, allows to conjugate these two reductions and decrease by two the order of the initial system (the degree of freedom is eliminated). Furthermore, the reduced problem remains posed in variational terms.} 

The dynamical function~\eqref{Routhian} was called \emph{modified Lagrangian} by Routh \cite{Routh1877}. We shall rather call it a \emph{reduced Lagrangian} or a \emph{Routhian function} \cite{Pars,Rutherford,MarsdenRatiu}. To the best of our knowledge, the first use of the letter $R$ to designate this function in honour of Routh is to be found in \textcite{Whittaker}.\footnote{In his treatise, Routh points out afterwards that, whether $q^1,\dots,q^m$ were ignorable or not, if each occurrence of a $C_i$ is replaced by $p_i$ in the right-hand side of~\eqref{Routhian}, the resulting function of $t$, the $q^i$, the velocities $\dot q^{m+1},\dots,\dot q^n$, and the momenta $p_1,\dots,p_m$, constitutes a partial transformation into the Hamiltonian formulation: it behaves like a Lagrangian for the $m$ first degrees of freedom and like a Hamiltonian (though with an opposite sign) for the others. This more general point of view was privileged by Routh afterwards \cite{Routh1882}, the `modified Lagrangian' appearing as its corollary. In some textbooks \cite{Goldstein,Landau}, the Routhian is introduced as the function
\begin{align*}
R=p_1\dot q^1+\dots+p_m\dot q^m-L
\end{align*}
in order to recover the usual sign in the Hamiltonian canonical equations.
} One of the most famous application of the process is certainly the ignoration of the azimuthal motion in central force fields which `converts' the rotational kinetic energy into the centrifugal potential \cite{Lanczos,Rutherford}. Reversing the argument, one can wonder if a given potential energy appearing in the formulation of a mechanical problem is not, after all, only an `apparent fiction' emerging from some ignorated degrees of freedom. This is, in substance, the terms of an old question essentially discussed by Thomson and Hertz at the end of the 19th century (one can also cite Helmholtz who searched for an interpretation of heat as the resultant of a cyclic motion taking place inside the core of thermodynamical systems). While Routh certainly considered his reduction procedure as an helpful mechanical theorem, chiefly for the study of steady motions, Thomson \cite{Thomson1885,Thomson1888} and Hertz \cite{principles} were questioning through it the possibility of reducing the concept of `potential energy' to purely kinematical considerations in terms of concealed motions. More on the positions of these protagonists can be found in \textcite{Lutzen}.\footnote{Remarkably, the theory of Kaluza and (Oskar) Klein \cite{Kaluza1921,OKlein1926} is precisely the realisation of the program of Thomson and Hertz to electromagnetism. In this theory, the electromagnetic gauge field results from a cyclic motion taking place in a hidden fifth dimension which is topologically equivalent to a circle.}

In essence, the Routh reduction makes a bridge between equivalent variational formulations of a given problem. Another famous application of the procedure is notably the emergence of the historical \emph{least action principle} as the result of the ignoration of time in the extended Hamilton's one \cite{Murnaghan,Lanczos,Bazanski} that was reviewed in the previous section. Indeed, if $L$ does not depend on $t$, this variable is an ignorable coordinate of the extended Lagrangian $\mathscr L$. Consequently, $p_0=-H$ is a first integral. Let $-E$ be its actual value. Before being allowed to process to the ignoration of $q^0=t$, one must firstly verify that the equality
\begin{align}
\!\Theta(q^i,v^0,v^i)\coloneqq\frac{v^i}{v^0}\,\frac{\partial L}{\partial\dot q^i}\bigg(q^k,\frac{v^k}{v^0}\bigg)-L\bigg(q^k,\frac{v^k}{v^0}\bigg)=E\label{identity_E}
\end{align}
determines unambiguously $v^0$ as a function of the $q^i$, the $v^i$, and $E$. Since $\mathscr L$ is singular, this step, which is for example missing in \textcite{Lanczos}, cannot be overlooked. Taking the partial derivative of $\Theta$ with respect to $v^0$ gives
\begin{align*}
\frac{\partial\Theta}{\partial v^0}=\frac{v^iv^j}{(v^0)^3}\,\frac{\partial^2L}{\partial\dot q^i\partial\dot q^j}\bigg(q^k,\frac{v^k}{v^0}\bigg)=\frac{\dot q^i\dot q^j}{v^0}\,\frac{\partial^2L}{\partial\dot q^i\partial\dot q^j}(q^k,\dot q^k).
\end{align*}
But this is always nonzero (except eventually at some isolated instants where all the velocities vanish simultaneously) by the hypothesis made on the regularity of $L$. Thus, by the implicit function theorem \cite{Spivak}, the equality~\eqref{identity_E} effectively determines $v^0$ as a function $\varphi(q^i,v^i,E)$ and one can form the `extended Routhian function'
\begin{align}
\mathscr R(q^i,v^i)&=\bigg[L\bigg(q^i,\frac{v^i}{\varphi(q^k,v^k,E)}\bigg)+E\bigg]\,\varphi(q^i,v^i,E)\notag\\
&=v^i\,\frac{\partial L}{\partial\dot q^i}\bigg(q^i,\frac{v^i}{\varphi(q^k,v^k,E)}\bigg).\label{extended_Routhian}
\end{align}

The parameter $\sigma$ remains insignificant: any choice of $\sigma$ leads to the same functional form of the extended Routhian. That simple observation, in conjunction with the fact that $\mathscr R$ is a Lagrangian governing the $n$ degrees of freedom carried by the $q^i$, guarantees the homogeneity of degree 1 of $\mathscr R$ in the $v^i$. This property was established in \textcite{Bazanski} by explicit computations.\footnote{A quick inspection of~\eqref{extended_Routhian} shows that $\mathscr R$ is homogeneous of degree 1 in the $v^i$ iff $\varphi$ has the same property. The proof of the homogeneity of degree 1 of $\varphi$ was given in \textcite{Bazanski} on the basis of the homogeneity of degree 0 of $\Theta$. The demonstration is very simple but it may be rendered easier by avoiding any calculations. Indeed, let $\lambda$ be a constant coefficient. On the one hand, one has by the definition of $\varphi$:
\begin{align}
\Theta(q^i,\varphi(q^i,\lambda v^i,E),\lambda v^i)=E.\label{homogeneity1}
\end{align} 
On the other hand, one has by the homogeneity of $\Theta$:
\begin{align}
\Theta(q^i,\lambda \varphi(q^i, v^i,E),\lambda v^i)=\Theta(q^i,\varphi(q^i,v^i,E),v^i)=E.\label{homogeneity2}
\end{align} 
Equations~\eqref{homogeneity1} and~\eqref{homogeneity2} induce the equality
\begin{align*}
\Theta(q^i,\varphi(q^i,\lambda v^i,E),\lambda v^i)=\Theta(q^i,\lambda \varphi(q^i, v^i,E),\lambda v^i)
\end{align*}
from which is extracted the sought homogeneity of $\varphi$.}

One is left with the reduced variational principle obtained by the ignoration of time:
\begin{align}
\delta\int v^i\,\frac{\partial L}{\partial\dot q^i}\bigg(q^j,\frac{v^j}{\varphi(q^k,v^k,E)}\bigg)\,\rmd \sigma=0\label{time_ignoration}
\end{align}
for any variation of the $q^i$ vanishing at the extremities. When $\sigma$ is chosen to be the time $t$, one recovers the `least action principle' as defined in \textcite{Whittaker}. The latter generalizes the old one which was focused on time-independent natural problems and which is often summarized by \cite{Lanczos,Sommerfeld,Goldstein}
\begin{align*}
\delta\int T\,\rmd t=0,
\end{align*}
up to a removable factor 2. However, in these problems, the function $\varphi$ is easily determined: 
\begin{align*}
\varphi(q^i,\dot q^i,E)=\sqrt{\frac{g_{ij}\dot q^i\dot q^j}{2(E-V)}}\,,
\end{align*}
leading quite naturally to the Jacobi principle \cite{Jacobi1866}:
\begin{align}
\delta\int\sqrt{(E-V)g_{ij}\rmd q^i\rmd q^j}=0,\label{Jacobi_principle}
\end{align}
which amount to seeking the geodesics of the manifold with respect to the modified metric whose components are $h_{ij}=(E-V)g_{ij}$. It is easily extended to the more general Lagrangian~\eqref{model} by adding to the functional~\eqref{Jacobi_principle} the circulation of $A$. Jacobi's principle is also often taken as synonym for the least action principle \cite{Lamb,Appell}. It is, in a way, its achievement regarding natural systems: the time $t$ is completely eliminated and the problem is now posed in purely geometric terms. 

The equations of the trajectory are derived from Jacobi's principle after the introduction of a parameter. The most natural choice is the arclength because it is an intrinsic quantity. However, in our desire of reduction, we may chose one of the coordinates, say $q^1$. In this way, one obtains a problem with $n-1$ degrees of freedom. The other side of the coin is that it is no more autonomous (unless $q^1$ were ignorable) and, more serious still, the reduction is far from being intrinsic. It is nevertheless interesting to note that this reduction can be realized for any Lagrangian by choosing $q^1$ as parameter $\sigma$ in~\eqref{time_ignoration}. The Lagrangian thus obtained is
\begin{align*}
L'(q^1,q^2,\dots,q^n,\mathring q^2,\dots,\mathring q^n)=\mathring q^i\,\frac{\partial L}{\partial\dot q^i}\bigg(q^j,\frac{\mathring q^j}{\varphi(q^k,\mathring q^k,E)}\bigg),
\end{align*}
where the empty bullet symbolizes the total derivative with respect to the independent variable $q^1$ while $\mathring q^1$ is simply the number 1. One recovers here the theorem of Whittaker \cite{Whittaker1900,Whittaker} on the reduction of the degrees of freedom `by means of the energy-equation'. The latter can thus be added to the list of principles and theorems inferred from the application of Routh reduction procedure.

\subsection{Successive reductions}

Hitherto, we have only considered individual NPS. Since one such symmetry amounts to the existence of a cyclic coordinate, one can always use it to reduce by one the number of degrees of freedom through the ignoration process. Now, let us suppose that $\Phi_1$ and $\Phi_2$ are two NPS of the general problem discussed in section~\ref{sec:Noether}, with BH terms $f_1$ and $f_2$ respectively. Let $\boldsymbol\xi_1$ and $\boldsymbol\xi_2$ be their generators. By the compatibility~\eqref{compatibility_ev} between the prolongation and the linear structure of vector fields, it is clear that, for any constant $\lambda$, $\boldsymbol\xi_1+\lambda\boldsymbol\xi_2$ is again a generator of NPS, with BH term $f_1+\lambda f_2$. Then, by the compatibility~\eqref{compatibility_bracket} with the bracket, one finds easily that $[\boldsymbol\xi_1,\boldsymbol\xi_2]$ is also a generator of NPS, with BH term $\boldsymbol\xi_1(f_2)-\boldsymbol\xi_2(f_1)$. It proves that the set of NPS forms a Lie group. 

Suppose that $\boldsymbol\xi_1$ and $\boldsymbol\xi_2$ are independent. There exists an equivalent Lagrangian admitting both the invariances under $\Phi_1$ and $\Phi_2$ iff one can find a function $\Lambda(t,q^i)$ such that 
\begin{align}
f_1+\boldsymbol\xi_1(\Lambda)=0\quad\text{and}\quad f_2+\boldsymbol\xi_2(\Lambda)=0.\label{double_condition}
\end{align}
In addition to the existence of such a gauge, one will be able to convert the two invariances by the independence on two extended coordinates iff $\boldsymbol\xi_1$ and $\boldsymbol\xi_2$ commute. One sees that the possibility of converting two independent NPS into two ignorable coordinates are subjected to strong conditions. Suppose that $\boldsymbol\xi_1$ and $\boldsymbol\xi_2$ commute. Applying $\boldsymbol\xi_1$ on the right equality of~\eqref{double_condition}, $\boldsymbol\xi_2$ on the left one, and subtracting, one obtains the necessary condition $\boldsymbol\xi_1(f_2)-\boldsymbol\xi_2(f_1)=0$. By Poincar\'e's lemma on differential forms, this condition is also sufficient when $\mathcal E$ is two-dimensional. 

For the sake of illustration, consider the Lagrangian
\begin{align*}
L=\frac12\,\dot q^2+Fq
\end{align*}
which traditionally describes the rectilinear dynamics of a particle submitted to a uniform and time-independent force $F$. The problem, in its terms, is obviously invariant under time and space translations. These invariances manifest themselves by the NPS $\boldsymbol\xi_1=\partial_t$ and $\boldsymbol\xi_2=\partial_q$, with BH terms $f_1=0$ and $f_2=Ft$, respectively. Since $\boldsymbol\xi_1(f_2)-\boldsymbol\xi_2(f_1)=F$, it is impossible to find an equivalent Lagrangian admitting the two symmetries of the problem when $F\ne 0$. At most, we can work with a $t$-independent Lagrangian or a $q$-independent one. If, for some reason, we are more interested in the $q$-independence, it suffices to introduce the function $\Lambda=-Ftq$ verifying $f_2+\partial_q(Ftq)=0$ and to work with the equivalent Lagrangian
\begin{align*}
L'=L+\dot\Lambda=\frac12\,\dot q^2-Ft\dot q.
\end{align*}   

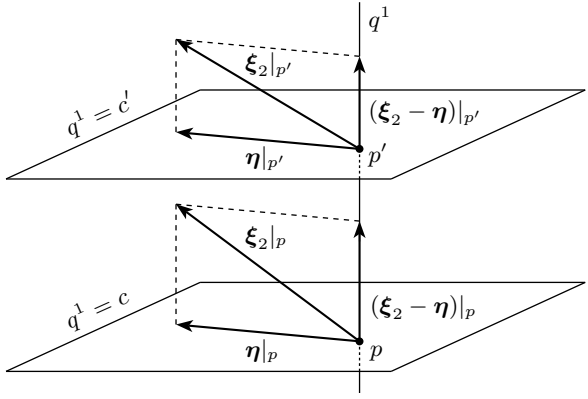
\begin{figure}
\psset{xunit=1.0cm,yunit=.8cm,algebraic=true,dotstyle=o,dotsize=3pt 0,linewidth=0.5pt,arrowsize=3pt 2,arrowinset=0.25}
\begin{center}
\begin{pspicture*}(0.7,0.7)(9,7.2)
\psline(6.08,1.06)(8.66,2.54)
\psline(6.08,1.06)(0.96,1.06)
\psline(0.96,1.06)(3.54,2.54)
\psline(3.54,2.54)(8.66,2.54)
\psline(6.08,4.26)(8.66,5.74)
\psline(6.08,4.26)(0.96,4.26)
\psline(0.96,4.26)(3.54,5.74)
\psline(3.54,5.74)(8.66,5.74)
\psline(5.66,0.7)(5.66,1.06)
\psline[linestyle=dashed,dash=1pt 1pt](5.66,1.06)(5.66,1.56)
\psline(5.66,3.22)(5.66,4.26)
\psline[linestyle=dashed,dash=1pt 1pt](5.66,4.26)(5.66,4.76)
\psline(5.66,6.3)(5.66,7.2)
\psline[linewidth=0.8pt]{->}(5.66,1.56)(3.22,1.84)
\psline[linewidth=0.8pt]{->}(5.66,4.76)(3.22,5.04)
\psline[linewidth=0.8pt]{->}(5.66,1.56)(3.22,3.84)
\psline[linewidth=0.8pt]{->}(5.66,1.56)(5.66,3.56)
\psline[linewidth=0.8pt]{->}(5.66,4.76)(5.66,6.3)
\psline[linewidth=0.8pt]{->}(5.66,4.76)(3.22,6.58)
\psline[linestyle=dashed,dash=2pt 2pt,linewidth=0.5pt](3.22,1.84)(3.22,3.84)
\psline[linestyle=dashed,dash=2pt 2pt,linewidth=0.5pt](3.22,5.04)(3.22,6.58)
\psline[linestyle=dashed,dash=2pt 2pt,linewidth=0.5pt](3.22,3.84)(5.66,3.56)
\psline[linestyle=dashed,dash=2pt 2pt,linewidth=0.5pt](3.22,6.58)(5.66,6.3)
\psdots[dotstyle=*](5.66,1.56)
\rput[bl](5.8,1.24){$p$}
\rput[bl](1.7,1.7){\rotatebox{24.2}{$q^1=c$}}
\rput[bl](5.8,1.93){$(\boldsymbol\xi_2-\boldsymbol\eta)|_{p}$}
\psdots[dotstyle=*](5.66,4.76)
\rput[bl](5.78,4.46){$p'$}
\rput[bl](1.7,4.9){\rotatebox{24.2}{$q^1=c'$}}
\rput[bl](5.78,5.15){$(\boldsymbol\xi_2-\boldsymbol\eta)|_{p'}$}
\rput[bl](4.14,1.2){$\boldsymbol\eta|_p$}
\rput[bl](4.14,3.1){$\boldsymbol\xi_2|_{p}$}
\rput[bl](4.14,4.4){$\boldsymbol\eta|_{p'}$}
\rput[bl](4.14,5.95){$\boldsymbol\xi_2|_{p'}$}
\rput[bl](5.78,6.7){$q^1$}
\end{pspicture*}
\end{center}
\caption{Schematic representation of the foliation of $\mathcal E$ induced by the vector field $\boldsymbol\xi_1$ through the introduction of an adapted system $\{q^\mu\}$ such that $\boldsymbol\xi_1=\partial_1$. Vertical lines are integral curves of $\boldsymbol\xi_1$ while horizontal subspaces are leaves of the foliation. The reduced space of events results from the identification between points in the same vertical line, leaf after leaf (here $p$ and $p'$ are identified). The vector field $\boldsymbol\xi_2$ can be quotientized iff its projection $\boldsymbol\eta$ into the leaves is invariant, as illustrated in the figure.}\label{fig:foliation}
\end{figure}

When the two independent NPS $\Phi_1$ and $\Phi_2$ verify the conditions leading to two ignorable coordinates, one can at once reduce the degrees of freedom by two thanks to Routh procedure. However, these conditions are \textit{a priori} too restrictive: to reduce the degrees of freedom by two, it suffices to be able to make two successive reductions by one. Performing a change of coordinates and gauge if necessary, one can suppose that $L$ is already a $\Phi_1$-invariant Lagrangian expressed in a system $\{t,q^i\}$ adapted to $\boldsymbol\xi_1=\partial_1$. The momentum $p_1$ being the first integral induced by $\Phi_1$, let $\varphi$ be the function such that
\begin{align*}
p_1=C\iff \dot q^1=\varphi(t,q^2,\dots,q^n,\dot q^2,\dots,\dot q^n,C).
\end{align*}
Then, by Routh procedure, one obtains the reduced Lagrangian
\begin{align*}
R(t,q^2,\dots,q^n,\dot q^2,\dots,\dot q^n)=L-C\dot q^1,
\end{align*}
each occurrence of $\dot q^1$ in the right-hand side being implicitly understood as the function $\varphi$. Locally, the reduced space of events is obtained by identifying all points of a same integral curve of $\Phi_1$. To put it another way, points differing only by the value of their coordinate $q^1$ are no more distinguished. The adapted coordinate system $\{t,q^i\}$ realizes locally a foliation of $\mathcal E$ into leaves of equation $q^1=\text{cst}$ (see figure~\ref{fig:foliation}). At each point $p$ the vector $\boldsymbol\xi_2$ can be `projected' into the leaf passing through $p$ to give the vector
\begin{align*}
\boldsymbol\eta=\sum_{\mu\ne 1}\xi^\mu_2\partial_\mu=\boldsymbol\xi_2-\xi_2^1\partial_1\,.
\end{align*}
Consider, now, an integral curve of $\boldsymbol\xi_1$. It is transverse to the family of leaves and $\boldsymbol\xi_2$ defines a `projected vector field' $\boldsymbol\eta$ along it. If, and only if, its components do not depend on the point along the integral curve then $\boldsymbol\eta$ can be `quotientized' into a genuine vector field over the reduced space of events. In other words, its components must not depend on $q^1$. It is quite simple to verify that this condition amounts intrinsically to a commutation rule of the form
\begin{align}
\big[\boldsymbol\xi_1,\boldsymbol\xi_2\big]=g\,\boldsymbol\xi_1,\label{commutation_rule}
\end{align}
where $g$ is some scalar field over $\mathcal E$.

We now have to answer to the following question: under this hypothesis, is $\boldsymbol\eta$ the generator of an NPS $\Phi_2'$ of the reduced variational problem? If so, it will allow us to decrease by one a second time the number of degrees of freedom. The prolongation of $\boldsymbol\eta$ with respect to the reduced space of events is
\begin{align*}
\boldsymbol\eta^{[1]}=\boldsymbol\eta+\sum_{i>1}\big(\dot\xi_2^i-\dot q^i\dot\xi_2^0\big)\frac{\partial}{\partial\dot q^i}\,,
\end{align*}
and one has
\begin{align*}
\boldsymbol\eta^{[1]}(R)&=\Big[\boldsymbol\xi_2^{[1]}(L)-\big(\dot\xi^1_2-\dot q^1\dot\xi_2^0\big)p_1\Big]_{\dot q^1=\varphi}\\
&=\bigg[\frac{\rmd}{\rmd t}\big(f_2-C\xi^1_2\big)\bigg]_{\dot q^1=\varphi}-\dot\xi_2^0 R\,.
\end{align*}
If $f_2-C\xi_2^1$ do not depend on $q^1$ then
\begin{align*}
\boldsymbol\eta^{[1]}(R)=\frac{\rmd}{\rmd t}\big(f_2-C\xi_2^1\big)-\dot\xi_2^0R
\end{align*}
and $\boldsymbol\eta$ is the generator of an NPS of the reduced problem. There are two possibilities: (i) either $f_2$ or $\xi_2^1$ depends on $q^1$ and $C$ has precisely the value such that $f_2-C\xi_2^1$ do not depend on $q^1$, (ii) neither $f_2$ nor $\xi_2^1$ depends on $q^1$ whereas $C$ is arbitrary. The first situation is non generic and will be put aside. One concludes that $f_2$ as well as all the components of $\boldsymbol\xi_2$ must not depend on $q^1$. But, recalling that $f_1=0$, it amounts to both the intrinsic conditions
\begin{align*}
\big[\boldsymbol\xi_1,\boldsymbol\xi_2\big]=0\quad\text{and}\quad \boldsymbol\xi_1(f_2)-\boldsymbol\xi_2(f_1)=0
\end{align*}
which put us back to the simultaneous reduction previously discussed. The first equality is the reason why Routh reduction can only be repeated when the group of NPS is Abelian, and the second one constitutes a further restriction. As a final remark, let us mention that a generalization of Routh procedure to the non-Abelian case exists --- although the reduced variational principle is no more of the Hamilton type --- and was achieved only recently \cite{Marsden1993}.

\section{Concluding remarks}

In writing this paper, we aimed to lay out the most significant issues regarding Noether's theory in classical mechanics, without hiding a certain aesthetic bias in the choice of the topics covered. We believe that fundamental physics cannot be contemplated without aesthetic motivations, if not emotions, and it is especially true concerning the idea of symmetry which is both a transcendental concept and a guiding principle, while being always more or less connected with our view of the world. 

In this year marking the centenary of Noether's article \textit{Invariante Variationsprobleme}, our work on the subject will be certainly one among many others and we hope that it will contribute, in its own way, to improve the understandings of her wonderful insight. 

\begin{acknowledgments}

The material of this paper is partly based on a conference given at the ninth annual colloquium `Cathy Dufour' which was devoted in 2016 to symmetries, invariances, and classifications. The author is indebted to Am\'elie Monjou for more than one decade of collaboration, C\'elia Krieger for her kind hospitality during the beginning of this work, \'Eric Adoul for his friendship, and he has a special though for H\'el\`ene Moraschetti. As usual, he is also grateful to the whole Statistical Physics Group, including Daniel Malterre, as well as to mathematicians and historians among whom must be mentioned Alain Genestier, Nicole Bardy-Panse, Fran\c cois Chargois, and Philippe Nabonnand. 

\end{acknowledgments}

\section*{Appendix}

\subsection*{`Generalized' symmetries}

In the body of the article, we focused our attention on point transformations, that is, on transformations of the events between themselves. But, since we are chiefly interested in evolutions, there is no reason to not considering more general transformations of them depending also on their velocities. Formally, it amounts to allow a dependency on the velocities of the components of the generators $\boldsymbol\xi$ whereas the prolongations formulas remain evidently unchanged.\footnote{Note that, in this case, the transformation of the position depends on the velocities, the transformation of the velocities depends on the second derivatives, etc. Hence, any $\mathscr C^{k}$ evolution ($k\geqslant 1$) is mapped into a $\mathscr C^{k-1}$ one.} The price to pay is obviously a more abstract geometric background which will not be discussed here and, worse still, the lost of the concept of adapted extended coordinates. Nonetheless, one will be able to give a more accurate definition of a Noether symmetry, closer to the original spirit of Noether and Bessel-Hagen.

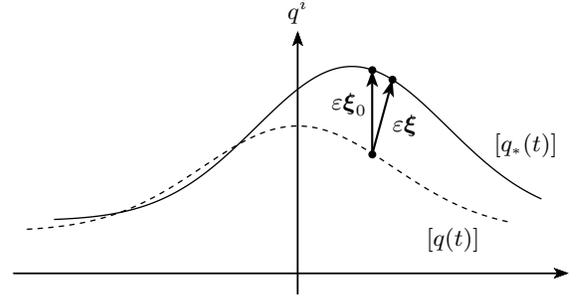
\begin{figure}
\begin{center}
\psset{xunit=1.8cm,yunit=1.4cm,algebraic=true,dimen=middle,dotstyle=o,dotsize=3pt 0,linewidth=0.5pt,arrowsize=3pt 2,arrowinset=0.25}
\begin{pspicture*}(-2.1,-0.6)(2.2,2.4)
\psaxes[labelFontSize=\scriptstyle,xAxis=true,yAxis=true,labels=none,ticks=none]{->}(-0.0,-.3)(-2.1,-0.5)(2.,2.)
\psplot[plotpoints=200,linestyle=dashed,dash=2pt 2pt]{-2}{1.55}{2.718281828459045^(-x^(2.0))+0.1}
\psplot[plotpoints=200]{-1.8}{1.8}{1.2*2.718281828459045^(-(x-0.4)^(2.0)+0.2)+0.2}
\psdots[dotstyle=*](0.55,0.83)
\psdots[dotstyle=*](0.7,1.54)
\psline[linewidth=.8pt]{->}(0.55,0.83)(0.7,1.54)
\psdots[dotstyle=*](.55,1.63)
\psline[linewidth=.8pt]{->}(0.55,0.83)(0.55,1.63)
\rput[bl](.25,1.2){$\varepsilon\boldsymbol\xi_0$}
\rput[bl](.95,-.1){{$[q(t)]$}}
\rput[bl](1.45,.8){{$[q_\sast(t)]$}}
\rput[bl](.7,1.){$\varepsilon\boldsymbol\xi$}
\uput{.1cm}[90](0.0,2){$q^i$}
\end{pspicture*}
\end{center} 
\caption{There is not an unique manner of transforming an evolution into another one. Transformations gather in equivalence classes. Here is shown the synchronous representative $\Phi_0$ of $\Phi$, generated by the generalized vector field $\boldsymbol\xi_0$ given by~\eqref{synchronous}.}\label{fig:vertical}
\end{figure}

The higher generality introduced here is not without redundancies. Let $[q(t)]$ be an evolution and assume that it is infinitesimally transformed into an evolution $[q_{\sast}(t)]$. They are infinitely close to each other with respect to some obvious notion of distance \cite{Gelfand}. In the limit $\varepsilon\to0$, the transformation tends to the identity and all the properties of $[q_{\sast}(t)]$ at the instant $t_{\sast}$ tend to the ones of $[q(t)]$ at the instant $t$. Since we are only concerned with the first order in $\varepsilon$, the difference between any two quantities infinitely close to each other multiplied by $\varepsilon$ will be neglected, as usual. The transformed evolution $[q_{\sast}(t)]$ is then readily obtained:
\begin{align*}
q_{\sast}^i(t)&=q_{\sast}^i(t_{\sast}-\varepsilon\,\tau)=q_{\sast}^i(t_{\sast})-\varepsilon\,\tau\dot q_{\sast}^i(t_{\sast})=q_{\sast}^i(t_{\sast})-\varepsilon\,\tau\dot{q}^i(t)\\
&=q^i(t)+\varepsilon(\xi^i-\dot q^i(t)\tau).
\end{align*}
This relation shows that all the transformations whose generators share the same \emph{characteristics} $\xi^i-\dot q^i\tau$ are equivalent in the sense that they map an evolution into a same other one, albeit in a different manner. In particular, the equivalence class of $\Phi$ contains an unique synchronous representative $\Phi_0$, \textit{videlicet}
\begin{align*}
(t,q^i(t))\longmapsto (t,q_{\sast}^i(t))=(t,q^i(t)+\varepsilon(\xi^i-\dot q^i(t)\tau)),
\end{align*} 
generated by (see figure~\ref{fig:vertical})
\begin{align}
\boldsymbol\xi_0=\big(\xi^i-\dot q^i\tau\big)\,\partial_i\,.\label{synchronous}
\end{align}

Now, one says that $\Phi$ is a \emph{Noether symmetry} of the variational problem if there exists a BH term $\mathscr F(t,q,\dot q)$ verifying~\eqref{variation} up to the first order in $\varepsilon$, for any evolution $[q(t)]$. Gathering $S[q(t)]$ and $S[q_\sast(t)]$ under a single integral as in~\ref{subsec:Noether}, one deduces that $\Phi$ is a Noether symmetry with BH term $\mathscr F$ iff~\eqref{RTidentity} is fulfilled, with $f(t,q^i,\dot q^i)$ the representative of $\mathscr F$. (In the special case where $\Phi$ is a point transformation, $\mathscr F$ cannot depend on the velocities and one recovers the context of~\ref{subsec:Noether}.) Except the end of~\ref{subsec:FI} where are considered the consequences of the two invariance issues, the discussion found in that subsection remains as it is and one infers from the symmetry the first integral~\eqref{INoether}. Moreover, the redundancy aforementioned gives rise to a third invariance issue in addition to those discussed in~\ref{subsec:FI}. 

\textit{(iii) Invariance under a change of representative}. Let us consider a transformation $\Phi'$ equivalent to $\Phi$ in the sense that its generator
\begin{align*}
\boldsymbol\xi'=\tau'\partial_t+\xi'^i\partial_i=\xi'^\mu\partial_\mu
\end{align*}
has the same characteristics than $\boldsymbol\xi$, i.e. is such that $\xi'^i-\dot q^i\tau'=\xi^i-\dot q^i\tau$. It is easily checked that
\begin{align*}
\text{RT}(L,\boldsymbol\xi,f)=\text{RT}(L,\boldsymbol\xi',f+(\tau'-\tau)L).
\end{align*}
Hence, $\boldsymbol\xi'$ is also a Noether symmetry, with BH term $f'=f+(\tau'-\tau)L$. It generates the same first integral
\begin{align*}
I=f-p_\mu\xi^\mu=f'-p_\mu\xi'^\mu.
\end{align*} 
The property of being a Noether symmetry or not is thus independent of the chosen representative of a transformation class. Furthermore, assuming that $\Phi$ is a Noether symmetry with BH term $f$, one sees that the transformation $\Phi_{\text N}$ generated by
\begin{align*}
\boldsymbol\xi_{\text N}=\bigg(\tau-\frac{f}{L}\bigg)\partial_t+\bigg(\xi^i-\dot q^i\,\frac{f}{L}\bigg)\partial_i
\end{align*}
is the only representative for which the symmetry is strict. Consequently, to each Noether symmetry corresponds a strict one giving rise to the same first integral, and even if uncountably many symmetries disappear by narrowing the study to strict invariance, none of their associated first integrals are lost. 

We shall mention that non point symmetries are generally hard to seek without making lucky ans\"atze, in contrast with point ones which can be found in an algorithmic way for most of them (see e.g. \textcite{Leone2015} for a case study about damped motions). Moreover, as was noticed by Noether, the interpretation of $I$ as an invariant of the symmetry is no more obvious and needs a careful analysis \cite{Sarlet_Cantrijn}.

\subsection*{The converse of Noether's theorem}

In this second part of the appendix, let us establish the converse of Noether's theorem \cite{Sarlet_Cantrijn,Leone2015}, viz.: to any first integral $I$ corresponds a Noether symmetry (and even an uncountable number of such symmetries). 

Let $I$ be a first integral. It is, by definition, a quantity depending on $t$, the $q^i$, and the $\dot q^i$, which is characterized by the vanishing of its total derivative along the motions:
\begin{align}
\frac{\rmd I}{\rmd t}\bigg|_{\{\mathsf E_i(L)=0\}}=0.\label{FItotal}
\end{align}
Since the Lagrangian is regular, the Euler-Lagrange equations can be put under the normal form $\ddot q^i=\Omega^i(t,q^k,\dot q^k)$ where, explicitly:
\begin{align*}
\Omega^i=\mathsf H^{ij}\bigg(\frac{\partial L}{\partial q^j}-\frac{\partial^2L}{\partial \dot q^j\partial t}-\dot q^k\,\frac{\partial^2L}{\partial\dot q^j\partial q^k}\bigg),
\end{align*}
with $(\mathsf H^{ij})$ the inverse of the Hessian matrix $(\mathsf H_{ij})$. The equivalence between the initial Euler-Lagrange equations and their normal form is rendered manifest by
\begin{align*}
\mathsf E_i(L)=\mathsf H_{ij}\big(\ddot q^{\,j}-\Omega^j\big).
\end{align*}
The introduction of the quantities $\Omega^i$ allows to replace~\eqref{FItotal} by the equivalent identity
\begin{align*}
\frac{\partial I}{\partial t}+\dot q^i\,\frac{\partial I}{\partial q^i}+\Omega^i\,\frac{\partial I}{\partial\dot q^i}=\frac{\rmd I}{\rmd t}+\mathsf H^{ij}\mathsf E_j(L)\,\frac{\partial I}{\partial\dot q^i}=0.
\end{align*}
Hence, $I$ is a first integral iff it verifies identically
\begin{align*}
\frac{\rmd I}{\rmd t}=-\mathsf H^{ij}\frac{\partial I}{\partial\dot q^j}\,\mathsf E_i(L).
\end{align*}
The quantities
\begin{align*}
\lambda^i=-\mathsf H^{ij}\frac{\partial I}{\partial\dot q^j}
\end{align*}
constitute a set of \emph{integrating factors} (or \emph{multipliers}) of the Euler-Lagrange equations associated with the first integral. Now, it is clear from~\eqref{RT2} that any transformation $\Phi$ generated by a vector field $\boldsymbol\xi$ having the $\lambda^i$ as characteristics will be a Noether symmetry with BH term $p_\mu\xi^\mu+I$. All of them form the class of Noether symmetries associated with the first integral $I$ whose synchronous representative $\Phi_0$ has for generator $\boldsymbol\xi_0=\mu^i\partial_i$ and for BH term $p_i\lambda^i+I$. The unique strict representative is thus generated by
\begin{align*}
\boldsymbol\xi_{\text N}=-\frac{p_i\lambda^i+I}{L}\,\partial_t+\bigg(\lambda^i-\dot q^i\,\frac{p_i\mu^i+I}{L}\bigg)\partial_i
\end{align*}
and coincides with the transformation that \textcite{Palmieri} introduced to establish the converse of Noether's theorem.\footnote{The integrating factors ($\lambda_i$) in \textcite{Palmieri} have an opposite sign than ours since these authors took $L^{(i)}=-\mathsf E_i(L)$ for the Lagrange expressions.} It can for example be used to find the symmetries associated with the conservation of the Laplace-Runge-Lenz vector, and compared with \textcite{Levy-Leblond1971}.

The existence of this strict representative is the reason why, even if one only deals with the strict invariance as in Noether's paper, each first integral corresponds nevertheless to a symmetry, and of course \emph{vice versa}. We must however mention that the converse exposed here does not correspond to the one which can be found in Noether's paper. Indeed, she proved the converse of the statement that each finite symmetry group of dimension $\rho$ generates $\rho$ linearly independent divergence relations.

\bibliography{Noether}

\end{document}